\documentclass[journal,twoside,web]{ieeecolor}
\usepackage{mathrsfs}
\let\labelindent\relax
\usepackage{generic}
\usepackage[switch]{lineno}
\usepackage{cite}
\usepackage{amsmath,amssymb,amsfonts}
\usepackage{graphicx}
\usepackage{algorithm}
\usepackage[dvipsnames]{xcolor}
\usepackage[colorlinks=true, allcolors=blue]{hyperref}
\usepackage{textcomp}
\usepackage[colaction]{multicol}
\usepackage{mathtools}

\usepackage{booktabs}
\usepackage{subcaption}
\usepackage{enumitem}
\usepackage{wrapfig}
\usepackage{algpseudocode}
\usepackage{tabularx}
\newtheorem{theorem}{Theorem}
\newtheorem{remark}{Remark}
\usepackage{lipsum} 
\newtheorem{lemma}{Lemma}

\newtheorem{proposition}{Proposition}
\usepackage{ragged2e}
\newtheorem{assumption}{Assumption}
\newtheorem{definition}{Definition}
\newtheorem{prob}{Problem}

\newcommand\Mycomb[2][^n]{\prescript{#1\mkern-0.5mu}{}C_{#2}}
\long\def\@makealgocaption#1#2{\vskip 2ex \small
  \hbox to \hsize{\parbox[t]{\hsize}{{\bfseries #1.} #2}}}

\def\BibTeX{{\rm B\kern-.05em{\sc i\kern-.025em b}\kern-.08em
    T\kern-.1667em\lower.7ex\hbox{E}\kern-.125emX}}
\begin{document}

\title{Minimum-Time Consensus of a Multi-Agent System under Fuel Constraints}
\author{Akansha Rautela, Deepak U. Patil, Ameer Mulla, and Indra Narayan Kar
\thanks{A. Rautela, D. U. Patil, and I. N. Kar are with the Dept. of Electrical
Engineering, IIT Delhi, India; A. Mulla is with the Dept. of Electrical
Engineering, IIT Dharwad, India (e-mail: Akansha.Rautela@ee.iitd.ac.in;
deepakpatil@ee.iitd.ac.in; ameer@iitdh.ac.in; ink@ee.iitd.ac.in).}
}
\maketitle
\begin{abstract}
This work addresses the problem of finding the \emph{minimum time} consensus point in the
state space for a set of $N$ identical double integrator agents with bounded input and
\emph{fixed fuel budget constraints}. The attainable set of each agent under these
constraints is computed, and its boundary is used to characterize the set of initial
conditions from which consensus is possible. The minimum time to consensus is the least
time at which the attainable sets of all agents intersect, and the corresponding consensus
state is their point of intersection. Using convexity of the attainable set and Helly's
theorem, this reduces to examining the $\Mycomb[N]{3}$ triplets of agents, for which a
closed-form expression for the minimum time and the consensus point is obtained. The
triplet computations are independent and are distributed evenly among the $N$ agents. The
overall minimum time and consensus point are given by the triplet with the greatest minimum
time, and each agent then computes its own control input locally. An example with $12$
agents is used to illustrate the results. 
\end{abstract}

\begin{IEEEkeywords}
Linear systems, Constrained control, Optimal control, Multi-agent systems.
\end{IEEEkeywords}

\section{Introduction}
\label{sec:introduction}
\IEEEPARstart{T}{he} study of a multi-agent system traces back to early efforts to reproduce, in computer animation, the collective behaviour observed in flocks of birds, swarms of insects and schools of fish \cite{boids}. A theoretical explanation of why such local, nearest-neighbour interaction rules give rise to coordinated group motion was later provided using graph-theoretic and matrix-theoretic tools \cite{jadbaieeetac2003}. These ideas were soon carried over to mobile robots and vehicles \cite{fax2004information}, and decentralized control of interconnected subsystems for objectives such as consensus, synchronization and formation has since been studied extensively \cite{4118472,ren2010distributed,inputconstraintsieeetac2018,safeconsensusbarrierfunctionieeetac2023,energystateconstraintsautomatica2024,flockingieeetac2023}, with applications in sensor networks \cite{cortes2005coordination}, power networks \cite{powernetworktac} and attitude alignment \cite{wang1999}.   
In consensus problems, a control law is sought such that each agent matches its state with that of the neighboring agent to reach a common agreement or point \cite{lin2003multi,dong2013leader}. { This article considers the problem of computing a consensus point that all agents can reach in \emph{minimum time} with \emph{bounded inputs} using a \emph{limited fuel budget}.} 

The speed of consensus can generally be improved by increasing graph connectivity (the Fiedler eigenvalue of the Laplacian) \cite{ren2010distributed}, but constraints on system resources, such as operational time and fuel or energy usage, must also be respected. Such situations arise in the rapid single-axis attitude alignment of multiple spacecraft with limited thruster fuel \cite{minimumswitchthruster,attitudecontrol}, and in surveillance, where multiple UAVs must quickly reassemble at a point or align in altitude within a limited fuel capacity \cite{fuelconstraintsinUAVs}.

We model the agents as double integrators, which capture a broad class of rotational and translational systems. For instance, in low-friction environments, single-axis spacecraft rotation is well approximated by a double integrator \cite{doubleintegrator}. The double integrator is also used for modeling the quadrotor, when the lower-level control is fast enough \cite{quaddoubleinteg,doubleinteg_xy}. Input, state, and energy constraints have recently been incorporated into consensus laws using model transformations and LMI-based approaches \cite{inputconstraintsieeetac2018,energystateconstraintsautomatica2024}, but optimal control objectives such as time and fuel remain largely unexplored in the multi-agent literature. Our earlier work addresses minimum-time consensus for double integrator agents \cite{doi:10.1080/00207179.2017.1285054,patil2019distributed} and, more recently, minimum-fuel consensus \cite{minfuelardpakink}.

A key difference between our method and existing approaches is the use of attainable and reachable sets, which naturally gives a characterization of the fundamental limitations on consensus in terms of initial conditions, time, and fuel. Our contribution is threefold: 1) we characterize the attainable set under a fixed fuel budget and bounded input, 2) using its structure, we characterize the set of initial states from which consensus is possible (Theorem \ref{thm:feas}) and compute the region in state-space where consensus occurs (Lemma \ref{lemma:consensuspossible}), 3) assuming the communication topology is a connected graph, we give Algorithms \ref{alg:tripassign} and \ref{alg:inputs} to compute the minimum time to consensus and the corresponding consensus state in a distributed manner, with each agent using Algorithm \ref{alg:inputs} to locally compute the control input that brings it to the consensus state exactly at the required time.

\section{\label{sec:descrptionandpreliminaries} Problem Formulation and Preliminaries}

\subsection{Problem Formulation}
Consider a set of $N$ identical double integrator agents with dynamics given by the following state-space equations
\begin{equation}
\dot{\mathbf{x}}_i = A\mathbf{x}_i+ Bu_i, \quad \mathbf{x}_i(0)=\mathbf{x}_{i0}  \quad \text{for} \quad i = 1, \hdots , N \label{n-agents}
\end{equation}
where $A =
\begin{bmatrix}
0 & 1\\
0 & 0 
\end{bmatrix},
B = 
\begin{bmatrix}
   0\\
   1  
\end{bmatrix}.$
The system states are  $\mathbf{x}_i= \begin{bmatrix} x_i &\dot{x}_i\end{bmatrix}^{\top} \in \mathbb{R}^{2}$. The input $u_i(t)$ is bounded $|u_i(t)|\le 1$ almost everywhere (a.e.) for $t\ge0$. The functional associated with {agent $a_i$} that describes the fuel consumption at time $t_f$ is $F(u_i(t),t_f):= \int_{0}^{t_f} |u_i(t)| dt$ (which is also the $L_1$ norm of input \cite{nagahara2013maximum}). 
Let the maximum allowed fuel consumption be given as $\beta$. The \emph{fuel budget constraint} is then written as: 
 \begin{equation} 
  F(u_i(t),t_f) \leq \beta  \label{eq:fuelbudget}
 \end{equation}
 Let the set of admissible inputs in the interval $[0,t_f]$ be: 
{\small{ \begin{equation}U_\beta:=\{u\in L^{\infty}[0,t_f]\big| F(u(t),t_f)\le \beta, |u(t)|\le 1\mbox{ a.e.}\}\label{admissible}\end{equation}}}
 \begin{definition}
    A multi-agent system is said to achieve \emph{consensus} if for all $i,j \in \{1, \hdots, N\}$, where $i \neq j$, $\|\mathbf{x}_i(t)-\mathbf{x}_j(t)\| \to 0$ as $t \to {t'}$ and $\mathbf{x}_i(t) = \mathbf{x}_j(t)$ for all $t \geq {t'}>0$. The time ${t'}$ is called the time to consensus, and the point $\mathbf{x}_i({t'})=\overline{\mathbf{x}} \mbox{ for }i = 1, \hdots, N$ is the corresponding consensus point. The consensus is said to be achieved in finite time if ${t'}<\infty$.
\end{definition} 

{For the set of $N$ identical agents given by \eqref{n-agents}, our goal is to compute the \emph{minimum time to consensus} $\bar{t}_f$ and the corresponding \emph{consensus point} $\mathbf{\bar{x}} \in \mathbb{R}^2$ such that $\mathbf{x}_i(\bar{t}_f)=\mathbf{x}_j(\bar{t}_f)=\mathbf{\bar{x}}$ and $F(u_i(t),\bar{t}_f) \leq \beta$ for all $i,j \in \{1,2 \hdots, N\}$ with $i \neq j$}.
{
We address this problem using the attainable set.
\begin{definition}
The \emph{Attainable Set} $\mathcal{A}_i^\beta(t_f,\mathbf{x}_{i0})$ is the set of all the states that an agent $a_i$ can reach from $\mathbf{x}_{i0} \in \mathbb{R}^2$ using admissible control $u_i(t) \in U_\beta$ at time $t_f>0$. 
Thus, {\small      $\mathcal{A}_i^\beta(t_f,\mathbf{x}_{i0}):= \Big\{ e^{At_f}\mathbf{x}_{i0}+ \int_0^{t_f}\hspace{-0.1cm} e^{A(t_f-\tau)}Bu_i(\tau) d\tau| u_i(t) \in U_\beta \Big\}$ }
\end{definition}
{Note that consensus among agents $a_i$, $i=1,...,N$ is possible at a given time $\tau$ if and only if $\bigcap_{i=1}^N \mathcal{A}_i^{\beta}(\tau,\mathbf{x}_{i0}) \neq \emptyset$. Thus, in terms of attainable sets, the problem is as follows. \begin{prob}\label{prob2}
  Compute $\bar{t}_f:=  \min\limits_{\bigcap_{i=1}^N \mathcal{A}_i^{\beta}(\tau,\mathbf{x}_{i0}) \neq \emptyset } \tau$, the corresponding $\mathbf{\bar{{x}}}\in \bigcap_{i=1}^N \mathcal{A}_i^{\beta}(\bar{t}_f,\mathbf{x}_{i0})$  and the control inputs which transfer agent $a_i$'s state from $\mathbf{x}_{i0}$ to $\mathbf{\bar{{x}}}$ at time $\bar{t}_f$.
\end{prob}
 {  \begin{remark}
  We consider identical fuel budgets for all agents for the sake of brevity. However, the presented approach is applicable for non-identical fuel budgets as well.
\end{remark}}

\subsection{Preliminaries}

Before characterizing attainable sets, we need the following notion of reachable set.
\begin{definition}
The \emph{Reachable Set}, $\mathcal{R}_i^\beta(t_f,\mathbf{0})$, is the set of initial conditions from which an agent $a_i$ can reach the origin $\mathbf{0}\in \mathbb{R}^2$ using admissible control $u_i(t) \in U_{\beta}$ in time $t_f>0$. 
Thus, \small     $\mathcal{R}_i^\beta(t_f,\mathbf{0}):= \Big\{\int_0^{t_f}\hspace{-0.1cm} e^{-A\tau}Bu_i(\tau) d\tau| u_i(t) \in U_{\beta} \Big\}$
    \normalsize
\end{definition}
 Note that $\mathcal{A}_i^\beta(t_f,\mathbf{x}_{i0})$ is obtained from $\mathcal{R}_i^\beta(t_f,\mathbf{0})$ by using $\mathcal{A}_i^\beta(t_f,\mathbf{x}_{i0}) = e^{At_f}(\mathbf{x}_{i0}+\mathcal{R}_i^\beta(t_f,\mathbf{0}))$ \cite{fashoro1992reachability}. {The points on the boundary of the set $\mathcal{R}_i^\beta(t_f,\mathbf{0})$ are
those initial conditions from which the \emph{minimum} fuel required to reach the
origin at time $t_f$ equals $\beta$. The interior points of $\mathcal{R}_i^\beta(t_f,\mathbf{0})$ require strictly less fuel than $\beta$.
Characterizing this boundary is therefore a minimum-fuel problem with fixed $t_f$
 whose solutions are given by the Lemma \ref{thm:bang-off-bang}. 
}} 
{
\begin{lemma}\label{thm:bang-off-bang}  \cite{athans2013optimal},\cite[Theorem 3.2]{Hjek1979L1optimizationIL}
    For the system given by \eqref{n-agents}, control ${u}(t)$ that steers state-trajectory $\mathbf{x}_i(t)$ of agent $a_i$ from an initial condition $\mathbf{x}_{i0} \in \mathbb{R}^2$ to the {  origin} with \emph{minimum} fuel i.e., $F(u(t),t_f)$ and in \emph{finite} time $t_f$ is of the form 
\begin{equation}
 \small  {u}(t)=\begin{cases}
    \gamma_1,  & \text{$t \in [0,t_1)$}\\
    0, & \text{$t \in [t_1,t_2)$} \qquad  0 \leq t_1 \leq t_2 \leq t_f< \infty\\
    \gamma_2  & \text{$t \in [t_2,t_f]$}\\
  \end{cases}\label{twoswitch}
\end{equation}
where $\gamma_j \in \{+1,-1\}$ for $j=1,2$.  
\end{lemma}
We denote the input of the form given in equation \eqref{twoswitch} by a sequence $(\gamma_1,0,\gamma_2)$. From Lemma \ref{thm:bang-off-bang}, the possible fuel optimal control sequences are therefore $s_1=(+1,0,-1)$, {$s_2 = (+1,0,+1)$}, $s_3=(-1,0,+1)$ and {$s_4=(-1,0,-1)$}.} These control sequences include a zero level, indicating that the control is consuming no fuel for a finite duration. {The fuel consumption for all sequences up to time $t_f$ is then $F(u(t),t_f)=t_f-t_2+t_1$. { Let
$\hat{U}_\beta:=\{u(t) \mbox{ s.t. \eqref{twoswitch} and } F(u(t),t_f)\le \beta\}\label{admissible}$.
Further, let {\small     $\hat{\mathcal{R}}_i^\beta(t_f,\mathbf{0}):= \Big\{\int_0^{t_f}\hspace{-0.1cm} e^{-A\tau}Bu_i(\tau) d\tau| u_i(t) \in \hat{U}_{\beta} \Big\}$} 
\begin{lemma}\label{Rhat=r} 
For a fixed $t_f\ge 0$, $\hat{\mathcal{R}}_i^\beta(t_f,\mathbf{0}) = {\mathcal{R}}_i^\beta(t_f,\mathbf{0})$
\end{lemma}
\begin{proof}
Since $\hat{U}_\beta\subseteq U_\beta$, $\hat{\mathcal{R}}_i^\beta(t_f,\mathbf{0}) \subseteq {\mathcal{R}}_i^\beta(t_f,\mathbf{0})$. Next, consider a point $\mathbf{x}\in {\mathcal{R}}_i^\beta(t_f,\mathbf{0})$. Since $\mathbf{x}$ can be steered to the origin at time $t_f$ using an admissible control from $U_\beta$, there exists a control $u_i(t)$ transferring $\mathbf{x}$ to the origin at $t_f$ with minimum fuel $\beta^\ast\le \beta$. By Lemma \ref{thm:bang-off-bang}, $u_i(t)$ is of the form \eqref{twoswitch}, and it satisfies $F(u_i(t),t_f)=\beta^\ast\le\beta$. Thus, $u_i\in\hat{U}_\beta$ and $\mathbf{x}\in \hat{\mathcal{R}}_i^\beta(t_f,\mathbf{0})$. Therefore, ${\mathcal{R}}_i^\beta(t_f,\mathbf{0})\subseteq\hat{\mathcal{R}}_i^\beta(t_f,\mathbf{0})$ and hence $\hat{\mathcal{R}}_i^\beta(t_f,\mathbf{0})={\mathcal{R}}_i^\beta(t_f,\mathbf{0})$.
\end{proof}
}

{  Since $\mathcal{A}_i^\beta(t_f,\mathbf{x}_{i0}) = e^{At_f}(\mathbf{x}_{i0}+\mathcal{R}_i^\beta(t_f,\mathbf{0}))$, we also use the fuel optimal control sequences to characterize the attainable set  $\mathcal{A}_i^\beta(t_f,\mathbf{x}_{i0})$.  }
For characterizing the boundary of the set $\mathcal{A}_i^\beta(t_f,\mathbf{x}_{i0})$, we need {$F(u_i(t),t_f) = \beta$ which signifies} the full utilization of the fuel budget. From Lemma \ref{Rhat=r}, we have
{\small 
$\mathcal{A}_i^\beta(t_f,\mathbf{x}_{i0})= \left\{ e^{At_f}\mathbf{x}_{i0}+\int_0^{t_f}\hspace{-0.2cm} e^{A(t_f-\tau)}Bu_i(\tau) d\tau| u_i(t) \in \hat{U}_\beta \right\}$
}
For a fixed $\mathbf{x}_{i0}$ and $t_f$, we have $\mathcal{A}_i^\beta(t_f,\mathbf{x}_{i0}) = e^{At_f} \mathbf{x}_{i0} + \mathcal{A}_i^\beta(t_f,\mathbf{0})$.
\begin{lemma}\label{convexity}
The set \(\mathcal{A}_i^\beta(t_f, \mathbf{x}_{i0})\) is
{ compact convex with non-empty interior}.
\end{lemma}
\begin{proof}
Consider two points $\mathbf{x}_1, \mathbf{x}_2 \in \mathcal{A}_i^\beta(t_f,\mathbf{x}_{i0})$. Then, $\mathbf{x}_1 = e^{At_f}\mathbf{x}_{i0} + \int_0^{t_f} e^{A(t_f-\tau)}Bu_1(\tau)\, d\tau$ with $|u_1(t)| \leq 1$, $F(u_1(t),t_f) \leq \beta$ and $\mathbf{x}_2 = e^{At_f}\mathbf{x}_{i0} + \int_0^{t_f} e^{A(t_f-\tau)}Bu_2(\tau) \, d\tau$ with $|u_2(t)| \leq 1$, $F(u_2(t),t_f) \leq \beta$.
Now, the convex combination $\mathbf{x}_\lambda = \lambda \mathbf{x}_1 + (1-\lambda) \mathbf{x}_2$ for $\lambda \in [0, 1]$ is given by
$\mathbf{x}_\lambda = e^{At_f}\mathbf{x}_{i0} + \int_0^{t_f} e^{A(t_f-\tau)}B \left(\lambda u_1(\tau) + (1-\lambda) u_2(\tau)\right) \, d\tau$.
Letting $u_\lambda(\tau) = \lambda u_1(\tau) + (1-\lambda) u_2(\tau)$ we note
$|u_\lambda(t)| \leq 1$
and
$F(u_\lambda(t), t_f) \leq \lambda F(u_1(t), t_f) + (1-\lambda) F(u_2(t), t_f) \leq \beta$.
Thus, $\mathbf{x}_\lambda \in \mathcal{A}_i^\beta(t_f,\mathbf{x}_{i0})$ and $\mathcal{A}_i^\beta(t_f,\mathbf{x}_{i0})$ is convex. 

{ The set $\mathcal{K}:=\{(t_1,t_2)\in\mathbb{R}^2|\ 0\le t_1\le t_2\le t_f,\ t_1+t_f-t_2\le\beta\}$
is compact with non-emty interior. For fixed $\gamma_i\in\{+1,-1\}$ for $i=1,2$, the function
{\small  $\mathbf{f}_{(\gamma_1,\gamma_2)}(t_1,t_2):=e^{At_f}\mathbf{x}_{i0}
  +\Big(\gamma_1\!\int_0^{t_1}\!\!
  +\gamma_2\!\int_{t_2}^{t_f}\!\!\Big) e^{A(t_f-\tau)}B\,d\tau$} 
is continuous on $\mathcal{K}$. For \eqref{n-agents}, the Jacobian of the map, $\mathbf{f}_{(\gamma_1,\gamma_2)}(t_1,t_2):=e^{At_f}\mathbf{x}_{i0}
  +\Big(\gamma_1\!\int_0^{t_1}\!\!
  +\gamma_2\!\int_{t_2}^{t_f}\!\!\Big) e^{A(t_f-\tau)}B\,d\tau$ has a determinant $\gamma_1\gamma_2(t_1-t_2)$ which is non-zero everywhere on $\mathcal{K}$ except when $t_1=t_2$. Thus, any open ball in $\mathcal{K}$ maps to an open ball in the set $\mathcal{A}_i^\beta(t_f,\mathbf{x}_{i0})$. Thus, the image $\mathbf{f}_{(\gamma_1,\gamma_2)}(\mathcal{K})$ is a compact set with non-empty interior. From Lemmas \ref{thm:bang-off-bang} and \ref{Rhat=r}, the set
{\footnotesize $\mathcal{A}_i^\beta(t_f,\mathbf{x}_{i0})=\bigcup\limits_{(\gamma_1,\gamma_2)\in \{-1,1\}^2}\mathbf{f}_{(\gamma_1,\gamma_2)}(\mathcal{K})$} is compact with non-empty interior.}
\end{proof}
\subsubsection*{Unlimited fuel}
We get $U_\infty=\{u\in L^{\infty}[0,t_f]\big||u(t)|\le 1 \mbox{ a.e.}\}$ by relaxing the fuel budget constraint \eqref{eq:fuelbudget}. It is well known that the attainable set $\mathcal{A}_i^\infty(t_f,\mathbf{x}_{i0})$ is characterized by the bang-bang control sequences \cite{fashoro1992reachability}. The bang-bang control sequences are of the following form and have no duration of time in which the input is zero.
 \begin{equation}
 \small {u}(t)=\begin{cases}
    \gamma,  & \text{$t \in [0,t_1)$}\\
    -\gamma  & \text{$t \in [t_1,t_f]$} 
\end{cases}\label{noswitch}
\end{equation} where  $\gamma\in\{-1,+1\}$.
Let $\hat{U}_\infty:=\{u(t) \mbox{ s.t. \eqref{noswitch}}\}$, then
    {\small $\mathcal{A}_i^\infty(t_f,\mathbf{x}_{i0})\hspace{-0.1cm}= \hspace{-0.1cm}\left\{ e^{At_f}\mathbf{x}_{i0}+\hspace{-0.1cm}\int_0^{t_f}\hspace{-0.1cm} e^{A(t_f-\tau)}Bu_i(\tau) d\tau | u_i(t) \in \hat{U}_\infty \right\}$}
 { \begin{remark}
    Note that in case of unlimited fuel budget the minimum time to consensus is {\small$\bar{t}:=  \min\limits_{\bigcap_{i=1}^N \mathcal{A}_i^{\infty}(\tau,\mathbf{x}_{i0}) \neq \emptyset } \tau$} and the corresponding consensus state is $\mathbf{\bar{{x}}}\in \bigcap_{i=1}^N \mathcal{A}_i^{\infty}(\bar{t},\mathbf{x}_{i0})$. The computation of $\bar{t}$ and $\bar{\mathbf{x}}$ has been explored in our previous work \cite{doi:10.1080/00207179.2017.1285054}. Here, we additionally impose a constraint of the limited fuel budget. Consequently, $\bar{t}_f\ge \bar{t}$.
\end{remark}
{\begin{lemma}\label{lem:inclusion}
    For $t_f'\ge t_f$, $e^{A(t_f'-t_f)}\mathcal{A}_i^\beta(t_f,\mathbf{x}_{i0})\subseteq\mathcal{A}_i^\beta(t'_f,\mathbf{x}_{i0})$
\end{lemma}
\begin{proof}
    Let $t'_f\ge t_f$ and a point $\mathbf{x}\in \mathcal{A}_i^\beta(t_f,\mathbf{x}_{i0})$. Then $\mathbf{x} = e^{At_f}\mathbf{x}_{i0} + \int_0^{t_f} e^{A(t_f-\tau)}Bu_1(\tau)\, d\tau$ with $|u_1(t)| \leq 1$, $F(u_1(t),t_f) \leq \beta$. Now, $e^{A(t_f'-t_f)}\mathbf{x}=e^{At'_f}\mathbf{x}_{i0} + \int_0^{t'_f} e^{A(t'_f-\tau)}Bu_2(\tau)\, d\tau$ with $u_2(t)=u_1(t)$ for $t\in[0,t_f)$ and $u_2(t)\equiv0$ for $t\in[t_f,t_f']$. Thus, $|u_2(t)| \leq 1$ for $t\in[0,t_f']$, $F(u_2(t),t'_f) \leq \beta$.   Therefore, $e^{A(t_f'-t_f)}\mathbf{x}\in \mathcal{A}_i^\beta(t'_f,\mathbf{x}_{i0})$. Hence $e^{A(t_f'-t_f)}\mathcal{A}_i^\beta(t_f,\mathbf{x}_{i0})\subseteq\mathcal{A}_i^\beta(t'_f,\mathbf{x}_{i0})$.
\end{proof}}
\begin{lemma}\label{convergence}
Fix $t_f>0$ such that $\beta \ge t_f$, \(\mathcal{A}_i^\beta(t_f, \mathbf{x}_{i0})=\mathcal{A}_i^\infty(t_f, \mathbf{x}_{i0})\).
\end{lemma}
{
\begin{proof}
It always holds that $U_\beta\subseteq U_\infty$, and hence
$\mathcal{A}_i^\beta(t_f,\mathbf{x}_{i0})\subseteq\mathcal{A}_i^\infty(t_f,\mathbf{x}_{i0})$. For any $u_i\in U_\infty$, we get
$F(u_i(t),t_f)=\int_0^{t_f}|u_i(t)|\,dt\ \le \ t_f\ \le\ \beta $. Hence $u_i\in U_\beta$ and therefore $U_\infty\subseteq
U_\beta$. This gives
$\mathcal{A}_i^\infty(t_f,\mathbf{x}_{i0}) \subseteq\mathcal{A}_i^\beta(t_f,\mathbf{x}_{i0})$ and hence $\mathcal{A}_i^\beta(t_f,\mathbf{x}_{i0}) =\mathcal{A}_i^\infty(t_f,\mathbf{x}_{i0})$.
\end{proof}
}} 
Next, for $\beta\le t_f$, $\mathcal{A}_i^\beta(t_f,\mathbf{x}_{i0})$ is characterized by its boundary.

\subsection{Attainable Set Characterization} \label{sec:attainableset}
The states attained from $\mathbf{x}_{i0}$ at time $t_f$ using input \eqref{twoswitch} are  
\begin{equation}
\small   \mathbf{x}_i = e^{At_f}\mathbf{x}_{i0} + \left( {\gamma_1}\int_0^{t_{1}} + {\gamma_2}\int_{t_{2}}^{t_{f}} \right) e^{A(t_f-\tau)} B d\tau
\label{eq:attainablestates}
\end{equation}
where $\gamma_k\in \{-1,+1\}$ for $k=1,2$. Solving \eqref{eq:attainablestates} simultaneously with the constraint $t_f-t_2+t_1=\beta$ gives the boundary of the attainable set $\mathcal{A}_i^\beta(t_f, \mathbf{x}_{i0})$. We obtain four distinct parts of the boundary of the attainable set using input sequences $ s_k$, $ k=1,2,3,4$, respectively.

For the input sequence $s_1$, we solve the equations obtained by expanding \eqref{eq:attainablestates}. For notational purpose, we denote $t_1,t_2$ in this case as $t^{s_1}_1,t^{s_1}_2$. Solving equation \eqref{eq:attainablestates} simultaneously with constraint $t_f-t_2^{s_1}+t_1^{s_1}=\beta$ for the sequence $s_1$ gives
\begin{equation}\label{eq:switchtimess1}
\small  t^{s_1}_1 = \frac{\dot{x}_i + \beta - \dot{x}_{i0}}{2};~~ 
t^{s_1}_2 = \frac{\dot{x}_i - \beta + 2t_f - \dot{x}_{i0}}{2} 
\end{equation} along with an expression for the boundary points as $\Gamma^{{\beta},s_1}_{\mathbf{x_{i0}}}(\mathbf{x}_i,t_f)=0$, where {
{$\small \Gamma^{\beta,s_1}_{\mathbf{x}_{i0}}(\mathbf{x}_i,t_f)
  := x_i-x_{i0}-\frac{t_f}{2}\big(\dot{x}_i+\dot{x}_{i0}+\beta\big)
    +\frac{\beta^{2}+(\dot{x}_i-\dot{x}_{i0})^{2}}{4}.$ 
}}  Similarly, for sequence $s_3$ we get, 
 \begin{equation} \label{eq:switchtimess3}
\small      t^{s_3}_1 = \frac{-\dot{x}_i + \beta + \dot{x}_{i0}}{2};~~
        t^{s_3}_2 = \frac{-\dot{x}_i - \beta + 2t_f + \dot{x}_{i0}}{2}
\end{equation}
 along with an expression for the boundary points as $\Gamma^{{\beta},s_3}_{\mathbf{x_{i0}}}(\mathbf{x}_i,t_f)=0$ where
{ $\small \Gamma^{\beta,s_3}_{\mathbf{x}_{i0}}(\mathbf{x}_i,t_f)
  := x_i-x_{i0}-\frac{t_f}{2}\big(\dot{x}_i+\dot{x}_{i0}-\beta\big)
    -\frac{\beta^{2}+(\dot{x}_i-\dot{x}_{i0})^{2}}{4}.$
} The set of points $\mathbf{x}_i$ that satisfy equation $\Gamma^{\beta, s_1}_{\mathbf{x_{i0}}}(\mathbf{x}_i,t_f)=0$ or $ \Gamma^{\beta, s_3}_{\mathbf{x_{i0}}}(\mathbf{x}_i,t_f)=0$ forms a part of the boundary $\partial\mathcal{A}_i^\beta(t_f,\mathbf{x}_{i0})$ for these two control sequences. We denote these two parts as follows:
\begin{equation}  \label{eq:bds1}
 \small \partial{\mathcal{A}}_{i}^{\beta,s_1}:=\bigl\{ \mathbf{x}_i \in \mathbb{R}^2 ~\big|~  \Gamma^{{\beta},s_1}_{\mathbf{x_{i0}}}=0 \text{, } 0\leq t^{s_1}_1 \leq t^{s_1}_2 \le t_f \bigr\}
\end{equation}
\begin{equation} \label{eq:bds3} \small \partial{\mathcal{A}}_{i}^{\beta,s_3}:= \bigl\{ \mathbf{x}_i \in \mathbb{R}^2 ~\big|~ \Gamma^{{\beta},s_3}_{\mathbf{x_{i0}}}=0 \text{, } 0 \leq  t^{s_3}_1 \leq  t^{s_3}_2 \le t_f 
\bigr\}
\end{equation}
For the control sequence $s_2$,
\begin{equation}\label{eq:102-1}
 \small x_i = x_{i0} + \dot{x}_{i0}t_f +t_1^{s_2}t_f-t_2^{s_2}t_f -\frac{(t_1^{s_2})^2}{2} + \frac{(t_2^{s_2})^2}{2} + \frac{t_f^2}{2}
\end{equation}
 \begin{equation}\label{eq:102-2} \small \Gamma^{{\beta},s_2}_{\mathbf{x_{i0}}}(\mathbf{x}_i,t_f) :=\dot{x}_i - \dot{x}_{i0} - {\beta} = 0.
\end{equation}
On solving \eqref{eq:102-1}, \eqref{eq:102-2} and $t_f-t_2+t_1=\beta$, we get
{\small \begin{equation} \label{eq:switchtimess2}
   \small  \begin{aligned}
 t^{s_2}_1 &= \frac{-2x_i + 2x_{i0}+(\dot{x}_i-\dot{x}_{i0}-t_f)^2-t_f^2+2t_f\dot{x}_i}{2(\dot{x}_i-\dot{x}_{i0}-t_f)}
        \\
        t^{s_2}_2 &= \frac{-2x_i + 2x_{i0}-(\dot{x}_i-\dot{x}_{i0}-t_f)^2-t_f^2+2t_f\dot{x}_i}{2(\dot{x}_i-\dot{x}_{i0}-t_f)}
    \end{aligned}
\end{equation}}
Similarly, for $s_4$, we get: 
{{\small  \begin{equation}\label{eq:101-1}
x_i = x_{i0} + \dot{x}_{i0}t_f-t_1^{s_4}t_f + t_2^{s_4}t_f  +\frac{(t_1^{s_4})^2}{2}-\frac{(t_2^{s_4})^2}{2} - \frac{t_f^2}{2}
\end{equation}}
{\small \begin{equation}\label{eq:101-2}
\Gamma^{{\beta},s_4}_{\mathbf{x_{i0}}}(\mathbf{x}_i,t_f) := \dot{x}_i - \dot{x}_{i0} + {\beta} = 0.
\end{equation}}}
 {\small\begin{equation} \label{eq:switchtimess4}
  \small  \begin{aligned}
 t^{s_4}_1 &= \frac{-2x_i + 2x_{i0}-(\dot{x}_i-\dot{x}_{i0}+t_f)^2+t_f^2+2t_f\dot{x}_i}{2(\dot{x}_i-\dot{x}_{i0}+t_f)}
        \\
        t^{s_4}_2 &= \frac{-2x_i + 2x_{i0}+(\dot{x}_i-\dot{x}_{i0}+t_f)^2-t_f^2+2t_f\dot{x}_i}{2(\dot{x}_i-\dot{x}_{i0}+t_f)}
    \end{aligned}
\end{equation}}
From equations \eqref{eq:102-2} and \eqref{eq:101-2}, we see that this part of the boundary has a fixed $\dot{x}_i$ coordinate. Thus, 
{\small\begin{equation}\label{eq:bds2} \small \partial{\mathcal{A}}_{i}^{{\beta},s_2}:=\bigl\{ \mathbf{x}_{i}\in \mathbb{R}^2 ~\big|~ \Gamma^{{\beta},s_2}_{\mathbf{x_{i0}}} = 0 ,0 \le t_1^{s_2} \leq t_2^{s_2} \le t_f \bigr\}
\end{equation}}
{\small \begin{equation} \label{eq:bds4} \small \partial{\mathcal{A}}_{i}^{{\beta},s_4}:=\bigl\{ \mathbf{x}_{i}\in \mathbb{R}^2 ~\big|~ \Gamma^{{\beta},s_4}_{\mathbf{x_{i0}}} = 0 ,0 \le t_1^{s_4} \leq t_2^{s_4}\le t_f \bigr\}
 \end{equation}}
{ \begin{lemma} \label{lem:curve}
For $t_f\ge\beta$, $\bigcup\limits_{k=1}^4\partial\mathcal{A}_i^{\beta,s_k}$ is a simple closed curve
 \end{lemma}
 \begin{proof}
The boundary $\partial{\mathcal{A}}_{i}^{\beta,s_1}$ is a parabolic arc joining
$(x_{i0}+\dot{x}_{i0}t_f-\beta t_f+\frac{\beta^2}{2},\dot{x}_{i0}-\beta)$
with
$(x_{i0}+\dot{x}_{i0}t_f+\frac{\beta^2}{2},\dot{x}_{i0}+\beta)$.
Further, $\partial{\mathcal{A}}_{i}^{\beta,s_2}$ continues as a line segment from
$(x_{i0}+\dot{x}_{i0}t_f+\frac{\beta^2}{2},\dot{x}_{i0}+\beta)$
to
$(x_{i0}+\dot{x}_{i0}t_f+\beta t_f-\frac{\beta^2}{2},\dot{x}_{i0}+\beta)$.
Similarly, $\partial{\mathcal{A}}_{i}^{\beta,s_3}$ is again a parabolic arc joining
$(x_{i0}+\dot{x}_{i0}t_f+\beta t_f-\frac{\beta^2}{2},\dot{x}_{i0}+\beta)$
with
$(x_{i0}+\dot{x}_{i0}t_f-\frac{\beta^2}{2},\dot{x}_{i0}-\beta)$.
Finally, $\partial{\mathcal{A}}_{i}^{\beta,s_4}$ continues as a line segment joining
$(x_{i0}+\dot{x}_{i0}t_f-\frac{\beta^2}{2},\dot{x}_{i0}-\beta)$
and
$(x_{i0}+\dot{x}_{i0}t_f-\beta t_f+\frac{\beta^2}{2},\dot{x}_{i0}-\beta)$. For $t_f>\beta$, the two parabolic arcs $\partial{\mathcal{A}}_{i}^{\beta,s_1}$ and $\partial{\mathcal{A}}_{i}^{\beta,s_3}$ do not intersect because $\Gamma_{\mathbf{x}_{i0}}^{\beta,s_3}-\Gamma_{\mathbf{x}_{i0}}^{\beta,s_1} = \beta t_f - \frac{\beta^2+(\dot{x}_i-\dot{x}_{i0})^2}{2}>\beta(t_f-\beta)>0$ for $\dot{x}_i\in [\dot{x}_{i0}-\beta,\dot{x}_{i0}+\beta]$. Further, at $t_f=\beta$, the line segments $\partial{\mathcal{A}}_{i}^{\beta,s_2}$ and $\partial{\mathcal{A}}_{i}^{\beta,s_4}$ reduce into a point. Thus, for any $t_f\ge \beta$,  $\bigcup\limits_{k=1}^4\partial\mathcal{A}_i^{\beta,s_k}$ forms a simple closed curve (see Fig. \ref{fig1}).
\end{proof}}
}
\begin{figure}[h!]
\centering
\includegraphics[scale=0.28]{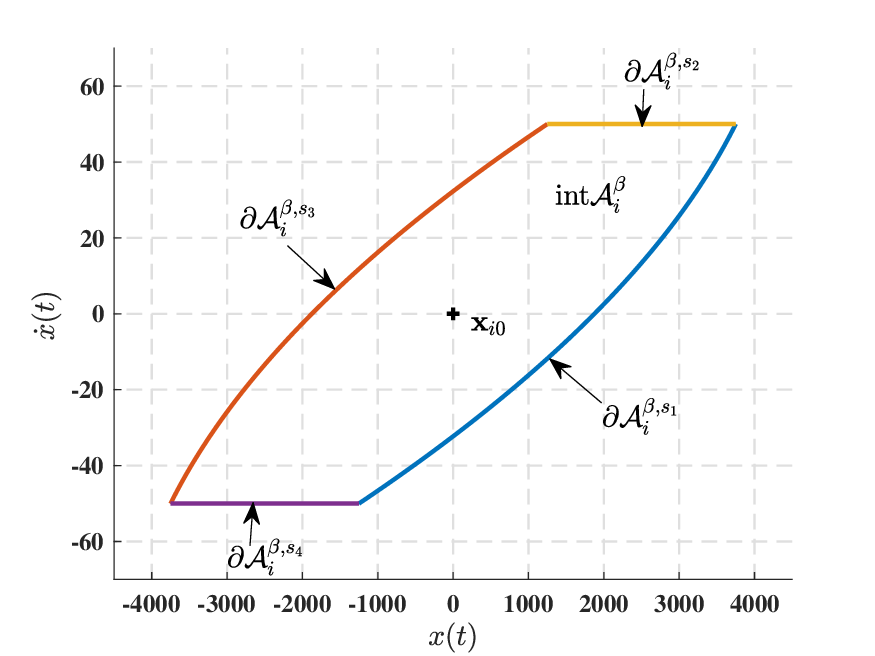}
\caption{\label{fig1}\footnotesize{ $\partial {\mathcal{A}}_i^{50}(t_f,\mathbf{x}_{i0})$ with $\mathbf{x}_{i0}=[0,0]^{\top}$}}
\end{figure}
{\begin{lemma}\label{lem:bd}
For $t_f\ge\beta$, the $\partial\mathcal{A}_i^\beta(t_f,\mathbf{x}_{i0})=\bigcup\limits_{k=1}^4\partial\mathcal{A}_i^{\beta,s_k}$ \end{lemma}
\begin{proof}
Recall $\mathcal{A}_i^\beta(t_f,\mathbf{x}_{i0}) = e^{At_f}(\mathbf{x}_{i0}+\mathcal{R}_i^\beta(t_f,\mathbf{0}))$ \cite{fashoro1992reachability}. Thus, the boundary of $\mathcal{R}_i^\beta(t_f,\mathbf{0})$ gives us the boundary of $\mathcal{A}_i^\beta(t_f,\mathbf{x}_{i0})$. Now, the points on the boundary of the set $\mathcal{R}_i^\beta(t_f,\mathbf{0})$ are those initial conditions which require $F(u_i(t),t_f)= \beta$ amount of fuel to reach the origin at time $t_f$. By Lemma \ref{thm:bang-off-bang}, the optimal control is necessarily of the form \eqref{twoswitch}. Thus, any point $\mathbf{x}\in \partial\mathcal{A}_i^\beta(t_f,\mathbf{x}_{i0})$ is obtained by using the form \eqref{twoswitch} and parameters $t_1$, $t_2$ satisfying $0\le t_1\le t_2\le t_f$ and $t_1+t_f-t_2=\beta$. Thus, it must lie in one of the sets given by \eqref{eq:bds1}, \eqref{eq:bds3}, \eqref{eq:bds2}, and \eqref{eq:bds4}, and hence $\partial\mathcal{A}_i^\beta(t_f,\mathbf{x}_{i0})\subseteq\bigcup\limits_{k=1}^4\partial\mathcal{A}_i^{\beta,s_k}.$ By Lemma \ref{convexity}, $\mathcal{A}_i^\beta(t_f,\mathbf{x}_{i0})$ is compact convex set with non-empty interior.  Hence $\partial\mathcal{A}_i^\beta$ is a simple closed curve. By Lemma \ref{lem:curve}, {\footnotesize$\bigcup\limits_{k=1}^4\partial\mathcal{A}_i^{\beta,s_k}$} is also a simple closed curve. Since a simple closed curve in $\mathbb{R}^2$ has no proper subset that is
itself a simple closed curve,  {\small $\partial\mathcal{A}_i^\beta(t_f,\mathbf{x}_{i0})=\bigcup\limits_{k=1}^4\partial\mathcal{A}_i^{\beta,s_k}$.}
\end{proof}
Lastly, for computing intersections, Helly's theorem stated next will be used.
\begin{lemma}\cite[Helly's Theorem]{gruber2007convex}\label{helly}
    Let $F$ be a finite family of at least $n+1$ convex sets on $\mathbb{R}^n$. If the intersection set of every $n+1$ members of $F$ is non-empty, then all members of $F$ have a non-empty intersection.
\end{lemma}
\section{{Consensus Region}}\label{sec:roc}
{ 
We define the following set in the $(x,\dot{x})$-plane.
{\small\begin{equation}\label{eqn:Xcb}
\mathcal{X}^{\beta}(\mathbf{x}_{i0}) :=\{(x,\dot{x})\in \mathbb{R}^2|\dot{x}_{i0}-\beta \leq \dot{x} \leq \dot{x}_{i0}+\beta\}
\end{equation}} 
Note that from Lemma \ref{lem:bd}, we get for any $t_f>0$, $\mathcal{A}_i^\beta(t_f, \mathbf{x}_{i0}) \subseteq \mathcal{X}^{\beta}(\mathbf{x}_{i0})$.
Let $\dot{x}_{\max}:= \max\limits_{i\in\{1,..,N\}}\,\dot{x}_{i0}$ and $\dot{x}_{\min}:= \min\limits_{i\in\{1,..,N\}}\,\dot{x}_{i0}$ denote the maximum and minimum $\dot{x}$ coordinates, respectively, among the initial conditions of all agents $a_i,~ i=1,..,N$.  Let $P:=\{i:\dot{x}_{i0}=\dot{x}_{\max}\}$, $Q:=\{j:\dot{x}_{j0}=\dot{x}_{\min}\}$, $x_{\max}:=\max\{x_{i0}:i\in P\}$, and
$x_{\min}:=\min\{x_{j0}:j\in Q\}$. Also, let the corresponding initial conditions be labeled as $\mathbf{x}_{\max,0}$ and $\mathbf{x}_{\min,0}$ respectively. Accordingly, denote the corresponding attainable sets as $\mathcal{A}_{\max}^{\beta}(t_f,\mathbf{x}_{\max,0})$ and $\mathcal{A}_{\min}^{\beta}(t_f,\mathbf{x}_{\min,0})$ respectively. Define the set $\mathcal{X}^{\beta}_c:=\bigcap\limits_{i=1}^N\mathcal{X}^{\beta}(\mathbf{x}_{i0})$. Thus, {\small \begin{equation}\label{consensuspoints}
 \mathcal{X}^{\beta}_c =\{(x,\dot{x})\in \mathbb{R}^2|\quad \dot{x}_{\max}-\beta \leq \dot{x} \leq \dot{x}_{\min}+\beta\}
\end{equation}}}
{\begin{lemma}\label{lemma:consensuspossible}
If consensus is possible, then the set $\mathcal{X}^{\beta}_c\ne \emptyset$. Conversely, if
$\mbox{int }\mathcal{X}^{\beta}_c \ne \emptyset$, then consensus is possible. Further, the
consensus point ${\mathbf{x}}_c\in\mathcal{X}^{\beta}_c$.
\end{lemma}
\begin{proof}
{If consensus occurs at ${\mathbf{x}_c}$ for some $t_f>0$, then
${\mathbf{x}_c}\in\mathcal{A}^{\beta}_i(t_f,\mathbf{x}_{i0})\subseteq\mathcal{X}^{\beta}(\mathbf{x}_{i0})$
for every agent $a_i$, hence
${\mathbf{x}_c}\in\mathcal{X}^{\beta}_c$. Therefore, $\mathcal{X}^{\beta}_c\neq\emptyset$. Next, if $\mbox{int }\mathcal{X}^{\beta}_c\ne \emptyset$ then there exist 
$\dot{{x}}_c\in(\dot{x}_{\max}-\beta,\dot{x}_{\min}+\beta)$ with
 $|\dot{{x}}_c-\dot{x}_{i0}|<\beta$ for all $i$. Substituting $\dot{x}_i=\dot{{x}}_c$ into
$\Gamma^{\beta,s_1}_{\mathbf{x}_{i0}}=0$ and $\Gamma^{\beta,s_3}_{\mathbf{x}_{i0}}=0$ and using Lemma \ref{lem:bd} for $t_f\ge \beta$, we get $x_i\in\mathcal{I}_i(t_f) = [l_i(t_f),w_i(t_f)]$ with
$l_i(t_f) := x_{i0}+\tfrac{1}{2}(\dot{x}_{i0}+\dot{{x}}_c)t_f-\frac{1}{2}\beta t_f+\tfrac{1}{4}(\beta^{2}+(\dot{{x}}_c-\dot{x}_{i0})^{2})$ and $w_i(t_f) := x_{i0}+\tfrac{1}{2}(\dot{x}_{i0}+\dot{{x}}_c)t_f+\frac{1}{2}\beta t_f-\tfrac{1}{4}(\beta^{2}+(\dot{{x}}_c-\dot{x}_{i0})^{2})$. Then, by Lemma \ref{helly}, $\cap_{i=1}^{N}\mathcal{I}_i(t_f)\ne \emptyset$ iff for all ordered pairs $1\le i,j\le N$, $ l_i(t_f)\le w_j(t_f)$. Now since $\mbox{int }\mathcal{X}_c^\beta\ne \emptyset$, $\dot{x}_{\max}-\dot{x}_{\min}<2\beta$. Thus, $(\beta-\frac{\dot{x}_{i0}-\dot{x}_{j0}}{2})>0$ for all ordered pairs $1\le i,j\le N$. Then, by rearranging, $l_i(t_f)\le w_j(t_f)$ we get $x_{i0}-x_{j0}+\frac{1}{2}\beta^{2}+\frac{1}{4}(\dot{{x}}_c-\dot{x}_{i0})^{2}+\frac{1}{4}(\dot{{x}}_c-\dot{x}_{j0})^{2}
\;\le\;
t_f(\beta-\frac{\dot{x}_{i0}-\dot{x}_{j0}}{2})$ which holds for all ordered pairs for a sufficiently large $t_f\ge\beta$.  Thus, $\cap_{i=1}^{N}\mathcal{I}_i(t_f)\ne \emptyset$. Picking ${x}_c\in\cap_{i=1}^{N}\mathcal{I}_i(t_f)$ gives ${\mathbf{x}_c}=[{x}_c,\dot{{x}}_c]^\top\in \mathcal{X}_c^\beta$  as a consensus point.
 }
\end{proof}This gives us the following theorem.
\begin{theorem}\label{thm:feas}
Consensus is possible if and
only if 
\begin{enumerate}
  \item[(i)] $\dot{x}_{\max}-\dot{x}_{\min}<2\beta$, or
  \item[(ii)] $\dot{x}_{\max}-\dot{x}_{\min}=2\beta$ and $x_{\min}-x_{\max}\ge\beta^{2}$.
\end{enumerate}
\end{theorem}
\begin{proof}
If consensus is possible then $\mathcal{X}_c^\beta \ne \emptyset$. Then either (a) $\mbox{int }\mathcal{X}_c^\beta \ne \emptyset$ or (b) $\mbox{int }\mathcal{X}_c^\beta = \emptyset$ but $\mathcal{X}_c^\beta \ne \emptyset$ holds. Now, (a) holds if and only if $\dot{x}_{\max}-\dot{x}_{\min}<2\beta$ giving necessity and sufficiency of (i) (by Lemma \ref{lemma:consensuspossible}). 

Next, if (b) holds then $\dot{x}_{\max}-\dot{x}_{\min}=2\beta$. From
\eqref{consensuspoints}, any consensus point $\mathbf{x}_c=[x_c,\dot{x}_c]^\top\in \mathcal{X}_c^{\beta}$ must be such that $\dot{{x}}_c=\tfrac{1}{2}(\dot{x}_{\max}+\dot{x}_{\min})$. Since $\dot{{x}}_c-\dot{x}_{\min}=\beta=-(\dot{{x}}_c-\dot{x}_{\max})$, $\mathcal{A}_{\min}^{\beta}$ and
$\mathcal{A}_{\max}^{\beta}$ touch on $\mathcal{X}_c^{\beta}$ only along 
$\partial\mathcal{A}_{\min}^{\beta,s_2}$ and
$\partial\mathcal{A}_{\max}^{\beta,s_4}$, respectively. From the proof of
Lemma~\ref{lem:curve}, $\partial\mathcal{A}_{\min}^{\beta,s_2}$ and
$\partial\mathcal{A}_{\max}^{\beta,s_4}$ are such that
$x_c\in[x_{\min}+\dot{x}_{\min}t_f+\frac{\beta^{2}}{2},\,
x_{\min}+\dot{x}_{\min}t_f+\beta t_f-\frac{\beta^{2}}{2}]$ and
$x_c\in[x_{\max}+\dot{x}_{\max}t_f-\beta t_f+\frac{\beta^{2}}{2},\,
x_{\max}+\dot{x}_{\max}t_f-\frac{\beta^{2}}{2}]$ respectively. Then, $\partial\mathcal{A}_{\min}^{\beta,s_2}\cap
\partial\mathcal{A}_{\max}^{\beta,s_4}\ne\emptyset$ if and only if
{\small
$\beta^{2}\;\le\;x_{\min}-x_{\max}\;\le\;2\beta t_f-\beta^{2}$.}
The
left inequality is independent of $t_f$, it must always hold. Hence, (ii) is necessary. 

Conversely, if (ii) holds, then  $\mathbf{x}_c=[x_c,\dot{x}_c]^\top\in \mathcal{X}_c^{\beta}$ must be such that $\dot{{x}}_c=\tfrac{1}{2}(\dot{x}_{\max}+\dot{x}_{\min})$. Note since $|\dot{x}_c-\dot{x}_{\max}|=\beta$, the agent $a_{\max}$ cannot attain $\dot{x}_c$ for any $t<\beta$. Hence consensus is not possible for $t_f<\beta$. For $t_f\ge \beta$ at $\dot{x}_i=\dot{x}_c$, we get $x_i\in\mathcal{I}_i(t_f)=[l_i(t_f),w_i(t_f)]$ where $l_i(t_f)$, and $w_i(t_f)$ are as defined in the proof of Lemma \ref{lemma:consensuspossible}. Thus, consensus at $t_f$ is possible iff
$\bigcap_{i=1}^{N}\mathcal{I}_i(t_f)\ne\emptyset$, i.e.\ iff $l_i(t_f)\le w_j(t_f)$
 for every ordered pair.  For pairs {\small $(i,j)\notin (P\times Q )$, $l_i(t_f)\le w_j(t_f)$} iff {\small $x_{i0}-x_{j0}+\frac{1}{2}\beta^{2}+\frac{1}{4}(\dot{{x}}_c-\dot{x}_{i0})^{2}+\frac{1}{4}(\dot{{x}}_c-\dot{x}_{j0})^{2}
\;\le\;
t_f(\beta-\frac{\dot{x}_{i0}-\dot{x}_{j0}}{2})$}, which holds iff {\small $t_f\ge \bar{t}_f:= \max\limits_{(i,j)\notin (P\times Q )}\frac{2(x_{i0}-x_{j0})+\beta^{2}+\frac{1}{2}(\dot{{x}}_c-\dot{x}_{i0})^{2}+\frac{1}{2}(\dot{{x}}_c-\dot{x}_{j0})^{2}}{(2\beta-{\dot{x}_{i0}+\dot{x}_{j0}})}$}. In particular, taking $(i,j)\notin P\times Q$ with $a_i=a_{\min}$ and $a_j=a_{\max}$ gives
$\bar{t}_f\ge\frac{x_{\min}-x_{\max}+\beta^{2}}{2\beta}\ge\beta$ by (ii). For pairs $(p,q)\in P\times Q$, we have
$\beta-\tfrac{1}{2}(\dot{x}_{p0}-\dot{x}_{q0})=0$, so
$l_p(t_f)\le w_q(t_f)$ reduces to $x_{p0}-x_{q0}+\beta^{2}\le0$, which is
independent of $t_f$. Further, it holds by (ii) since $x_{p0}\le x_{\max}$ and
$x_{q0}\ge x_{\min}$. These pairs therefore impose no lower bound on $t_f$. Thus, $\cap_{i=1}^{N}\mathcal{I}_i(t_f)\ne\emptyset$ iff $t_f\ge \bar{t_f}$. Hence, consensus is possible, and $\bar{t}_f$ is the minimum
time to consensus with the corresponding consensus point
$\bar{\mathbf{x}}=[\max_i l_i(\bar{t}_f),\tfrac{1}{2}(\dot{x}_{\max}+\dot{x}_{\min})]^{\top}$.
\end{proof}
\begin{remark}\label{rem:equality}
As the fuel budget is relaxed, i.e., as $\beta\to\infty$, any finite value of
$\dot{x}_{\max}-\dot{x}_{\min}$ eventually satisfies condition~(i) of
Theorem~\ref{thm:feas}. Condition~(ii) thus becomes irrelevant, and consensus is
possible for every initial configuration. Further,
$\mathcal{X}_c^{\beta}\to\mathbb{R}^{2}$ as $\beta\to\infty$.
\end{remark}
Note that if condition (ii) from Theorem \ref{thm:feas} holds, then the minimum time to consensus and the corresponding consensus point are already computed in the proof of Theorem \ref{thm:feas}. Thus, we next consider only cases in which the initial conditions of all agents satisfy condition (i) of Theorem \ref{thm:feas},  $\dot{x}_{\max}-\dot{x}_{\min}< 2\beta$. Here, the $\dot{x}$ coordinate of the consensus point is no longer fixed as in case (ii).}}

 %
\section{\label{sec2}Minimum time consensus}\label{sec:min.timeconsensus}
Recall the Problem \ref{prob2} whose goal is to compute $\bar{t}_f:=  \min\limits_{\bigcap_{i=1}^N \mathcal{A}_i^{\beta}(\tau,\mathbf{x}_{i0}) \neq \emptyset } \tau$. {This computation is intractable as the number of ways in which this intersection becomes non-empty grows exponentially as $O(2^N)$ with $N$}. Since each $\mathcal{A}_i^{\beta}$ is convex in $\mathbb{R}^{2}$, Helly's theorem
with $n=2$ reduces the problem to intersections of triplets.
\subsection{Minimum time consensus for a triplet of agents}\label{min.time.for.N=3}
{ Consider a triplet of agents $\{a_i,a_j,a_k\}$. Let the minimum time {\small $\bar{t}_{ijk} := \min\limits_{\bigcap\limits_{\alpha\in\{i,j,k\}}{\mathcal{A}}_{\alpha}^\beta(\tau, \mathbf{x}_{\alpha0})\ne\emptyset}{\tau}$}. Since, we need {\small $\bigcap\limits_{\alpha\in\{i,j,k\}}{\mathcal{A}}_{\alpha}^\beta(\tau, \mathbf{x}_{\alpha0})\ne\emptyset$}, for every pair of agents, $(a_i,a_j)$,$(a_j,a_k)$ and $(a_i,a_k)$, their attainable sets must intersect first. Let $\bar{t}_{ij}:=\min\limits_{\bigcap\limits_{\alpha\in\{i,j\}}{\mathcal{A}}_{\alpha}^\beta(\tau, \mathbf{x}_{\alpha0})\ne\emptyset}{\tau}$. Similarly, let $\bar{t}_{jk}$ and $\bar{t}_{ik}$ be the respective minimum time of intersection for the pair of agents $(a_j,a_k)$ and $(a_i,a_k)$. Note $\bar{t}_{ijk} \geq \text{max}\{\bar{t}_{ij} ,\bar{t}_{jk},\bar{t}_{ik}\}$. Let us assume without loss of generality $\bar{t}_{ij} = \text{max}\{\bar{t}_{ij} ,\bar{t}_{jk},\bar{t}_{ik}\}$ and the corresponding point of intersection $\bar{\mathbf{x}}_{ij}\in{\mathcal{A}}_{i}^\beta(\bar{t}_{ij}, \mathbf{x}_{i0})\cap\mathcal{A}_{j}^\beta(\bar{t}_{ij}, \mathbf{x}_{j0})$.} 
We now have two cases:

\textbf{Case (i):} If $\bar{\mathbf{x}}_{ij} \in \mathcal{A}_k^\beta(\bar{t}_{ij},\mathbf{x}_{k0})$ then, we get \emph{minimum time to consensus} as $\bar{t}_{ijk}=\bar{t}_{ij}$ and 
  the \emph{minimum time consensus state} $\bar{\mathbf{x}}_{ijk} = \bar{\mathbf{x}}_{ij}$. {An example scenario for case (i) is shown in Fig. \ref{fig21}.}

\textbf{Case (ii):} Otherwise we have $\bar{\mathbf{x}}_{ij} \notin \mathcal{A}_k^\beta(\bar{t}_{ij},\mathbf{x}_{k0})$. Here, even though for each pair of agents, the corresponding attainable sets intersect at time $\bar{t}_{ij} = \text{max}\{\bar{t}_{ij} ,\bar{t}_{jk},\bar{t}_{ik}\}$, but the intersection of all the three attainable sets is empty at $\bar{t}_{ij}$. 
Therefore, the first point of intersection will always occur on the boundaries of the attainable sets of the triplet of agents.
 To locate this point, we note that it must exist at some time $\bar{t}_{ijk}$ on the boundaries of all three attainable sets. Therefore, we must obtain all possible solutions $t_f$ that satisfy  
\begin{equation}\label{intersectionbd}
\partial {\mathcal{A}}_i^\beta(t_f, \mathbf{x}_{i0}) \cap \partial {\mathcal{A}}_j^\beta(t_f, \mathbf{x}_{j0}) \cap \partial{\mathcal{A}}_k^\beta(t_f, \mathbf{x}_{k0}) \neq \emptyset.
\end{equation}
Then $\bar{t}_{ijk}$ is chosen to be the least $t_f$ among all solutions. Further, $\mathbf{\bar{x}}_{ijk} \in \partial {\mathcal{A}}_i^\beta(\bar{t}_{ijk}, \mathbf{x}_{i0}) \cap \partial{\mathcal{A}}_j^\beta(\bar{t}_{ijk}, \mathbf{x}_{j0})\cap \partial{\mathcal{A}}_k^\beta(\bar{t}_{ijk}, \mathbf{x}_{k0})$ is the only point in the intersection.
{  Since $\mathbf{\bar{x}}_{ijk}$ lies on the boundary of attainable sets, control inputs of the form \eqref{twoswitch} are required to drive the states of all agents to $\mathbf{\bar{x}}_{ijk}$ at time $\bar{t}_{ijk}$.  Therefore, depending upon the structure of the input $\mathbf{\bar{x}}_{ijk}$ may lie on one of the four boundaries of $\partial{\mathcal{A}}_i^\beta(t_f, \mathbf{x}_{i0})$, $\partial{\mathcal{A}}_j^\beta(t_f, \mathbf{x}_{j0})$ and $\partial{\mathcal{A}}_k^\beta(t_f, \mathbf{x}_{k0})$. Solving \eqref{intersectionbd} requires computing solutions of a set of polynomial equations with inequality constraints simultaneously. 

Since there are $4$ parts to the boundaries, there are $64$ sets of three equations to be solved for computing $\mathbf{\bar{x}}_{ijk}$. {For discussing the validity of each of the 64 possibilities, we use $*$, $\bullet$, $\diamond$ as interchangeable placeholders for indices $i,~j$ and, $k$.} Since $\dot{x}_{\max}-\dot{x}_{\min}<2\beta$, out of these $64$ sets, we eliminate $18$ cases which only arise when $\dot{x}_{\max}-\dot{x}_{\min}=2\beta$.  These cases are the ones which arise when boundaries $\partial\mathcal{A}_{\ast}^{\beta,s_2}$ and $\partial\mathcal{A}_{\bullet}^{\beta,s_4}$ are involved simultaneously. For brevity, we have omitted the arguments while writing the sets.  
Since we assume $\dot{x}_{\max}-\dot{x}_{\min}<2\beta$,  $\bar{\mathbf{x}}_{ijk}$ will occur in one of the remaining $46$ cases.}

{Next, we argue that the possibilities where intersection of boundaries $\partial{\mathcal{A}}_{\bullet}^{\beta,s_l}$,  $\partial{\mathcal{A}}_{\diamond}^{\beta,s_l}$, with $l=1,3$ and $\partial{\mathcal{A}}_{\ast}^{\beta,s_r}$ with $r=2,4$ is involved simultaneously are also invalid. {Similarly, intersections involving sets $\partial{\mathcal{A}}_{\bullet}^{\beta,s_l}$,  $\partial{\mathcal{A}}_{\diamond}^{\beta,s_l}$, with $l=2,4$ and $\partial{\mathcal{A}}_{\ast}^{\beta,s_r}$ for $r = 1,3$ are also eliminated.}} There are in total 24 such possibilities. Note that in these cases even though $\partial{\mathcal{A}}_{\bullet}^{\beta,s_l} \cap \partial{\mathcal{A}}_{\ast}^{\beta,s_r} \cap \partial{\mathcal{A}}_{\diamond}^{\beta,s_l}$ is a singleton set, the ${\mathcal{A}}_{\bullet}^{\beta} \cap {\mathcal{A}}_{\ast}^{\beta} \cap {\mathcal{A}}_{\diamond}^{\beta}$ is not a singleton set and has non-zero area. This happens because the boundary segment $\partial{\mathcal{A}}_{\ast}^{\beta,s_r}$ in this case always has one of its end point in the interior of ${\mathcal{A}}_{\bullet}^{\beta}\cap {\mathcal{A}}_{\diamond}^{\beta}$.
 Since the ${\mathcal{A}}_{\bullet}^{\beta} \cap {\mathcal{A}}_{\ast}^{\beta} \cap {\mathcal{A}}_{\diamond}^{\beta}$ has non-zero area, $t_f$  can be reduced further till ${\mathcal{A}}_{\bullet}^{\beta} \cap {\mathcal{A}}_{\ast}^{\beta} \cap {\mathcal{A}}_{\diamond}^{\beta}$ shrinks to a singleton set. Similarly, four possibilities  $\partial{\mathcal{A}}_{\bullet}^{\beta,s_l} \cap \partial{\mathcal{A}}_{\ast}^{\beta,s_l} \cap \partial{\mathcal{A}}_{\diamond}^{\beta,s_l}$ for $l=1,2,3,4$ are eliminated. {Therefore, the minimum time consensus point $\bar{\mathbf{x}}_{ijk}$ occurs among the remaining following types, namely, 1) $\partial{\mathcal{A}}_{\bullet}^{\beta,s_1} \cap \partial{\mathcal{A}}_{\ast}^{\beta,s_1} \cap \partial{\mathcal{A}}_{\diamond}^{\beta,s_3}$, 2) $\partial{\mathcal{A}}_{\bullet}^{\beta,s_3} \cap \partial{\mathcal{A}}_{\ast}^{\beta,s_3} \cap \partial{\mathcal{A}}_{\diamond}^{\beta,s_1}$, 3) $\partial{\mathcal{A}}_{\bullet}^{\beta,s_1} \cap \partial{\mathcal{A}}_{\ast}^{\beta,s_2} \cap \partial{\mathcal{A}}_{\diamond}^{\beta,s_3}$, and 4) $\partial{\mathcal{A}}_{\bullet}^{\beta,s_1} \cap \partial{\mathcal{A}}_{\ast}^{\beta,s_3} \cap \partial{\mathcal{A}}_{\diamond}^{\beta,s_4}$ giving $18$ combinations on
assignments of $\bullet,*,\diamond$ to $i,j,k$.}  
{The time and state obtained by solving for each of these 18 cases is denoted by $t_{f,m}$ and $\hat{\mathbf{x}}_m$ for $m=1,\ldots,18$.} Among these $18$ possible cases, we discuss the solution of polynomial equations for two distinct possibilities -- 1) $\partial{\mathcal{A}}_{i}^{\beta,s_1} \cap \partial{\mathcal{A}}_{j}^{\beta,s_3} \cap \partial{\mathcal{A}}_{k}^{\beta,s_1} $  and 2) $\partial{\mathcal{A}}_{i}^{\beta,s_3} \cap \partial{\mathcal{A}}_{j}^{\beta,s_2} \cap \partial{\mathcal{A}}_{k}^{\beta,s_1}$. Remaining possibilities can be handled similarly.
\begin{figure}[h]
\begin{subfigure}{0.225\textwidth}
\includegraphics[scale=0.31]{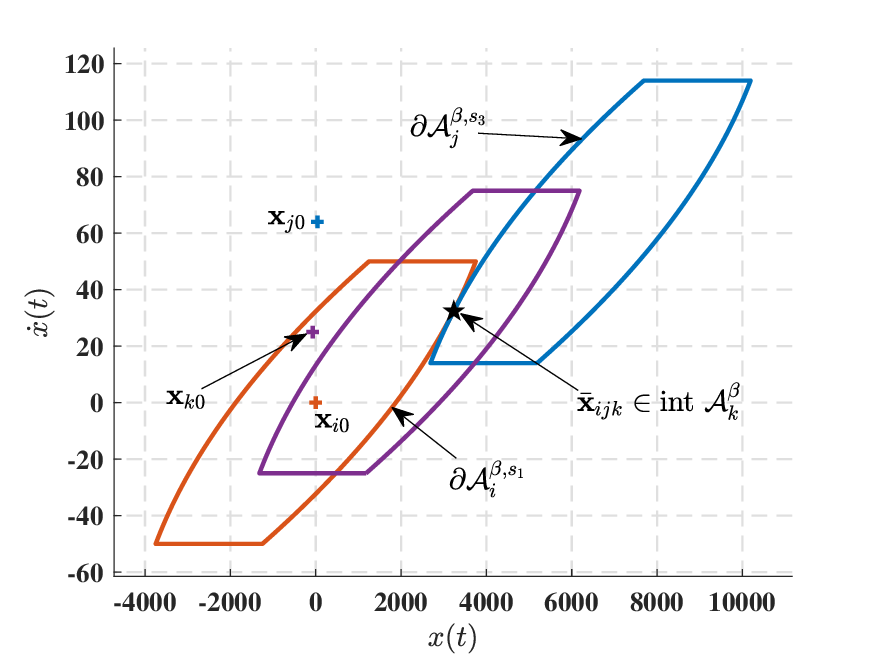}
    \caption{\label{fig21}{{Case (i)}}}
\end{subfigure}
\begin{subfigure}{0.222\textwidth}
\includegraphics[scale=0.25]{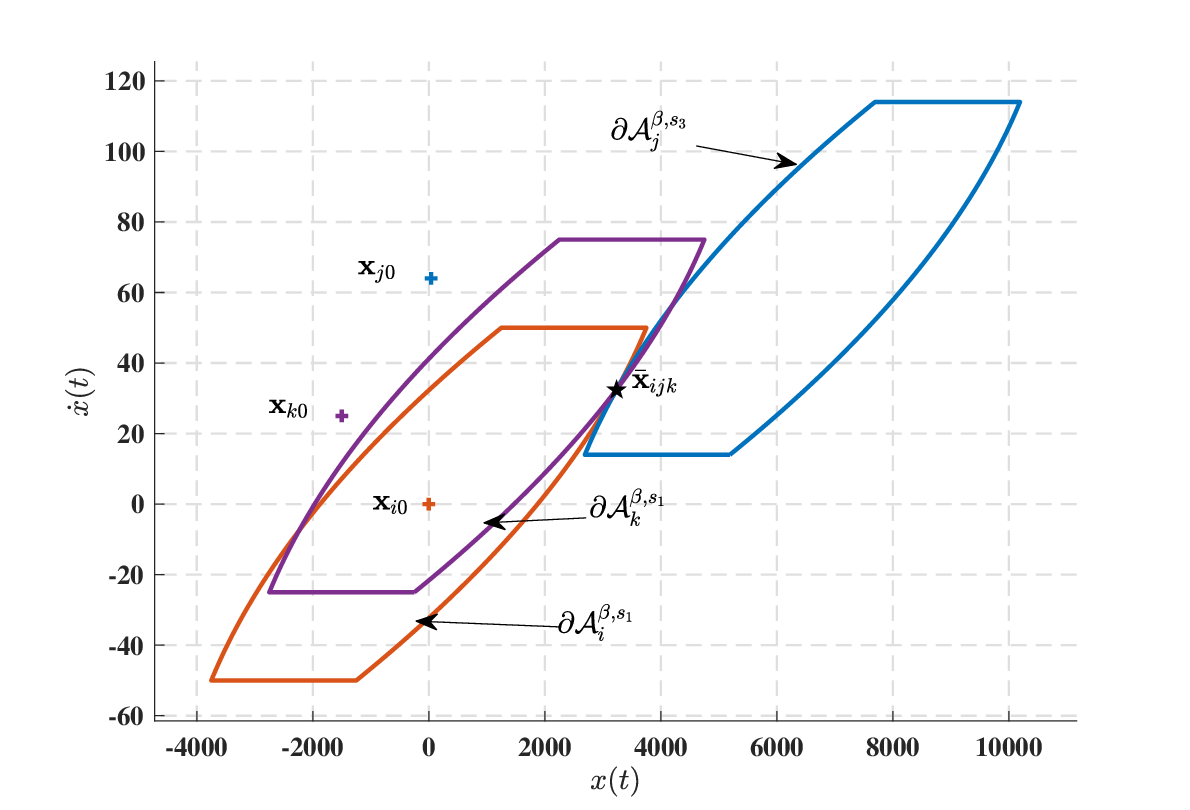}
\caption{\label{fig2}{Case (ii)}}
\end{subfigure}
\caption{Two cases for the intersection of attainable set for triplet $\{a_i,a_j,a_k\}$}
\end{figure}
\subsubsection{$\partial{\mathcal{A}}_{i}^{\beta,s_1} \cap \partial{\mathcal{A}}_{j}^{\beta,s_3} \cap \partial{\mathcal{A}}_{k}^{\beta,s_1} $  }\label{sec:boundaryequations1}

 In the boundary equations for $\partial {\mathcal{A}}_{i}^{\beta,s_1}$, $\partial {\mathcal{A}}_{j}^{\beta,s_3}$ and $\partial {\mathcal{A}}_{k}^{\beta,s_1}$, we substitute $\mathbf{x}_i = \mathbf{x}_j = \mathbf{x}_k = {\hat{\mathbf{x}}_1}$. As a result, we solve the three equations $\Gamma^{\beta,s_1}_{\mathbf{x}_{i0}}(\hat{\mathbf{x}}_1,t_{f,1}) = \Gamma^{\beta,s_3}_{\mathbf{x}_{j0}}(\hat{\mathbf{x}}_1,t_{f,1}) = \Gamma^{\beta,s_1}_{\mathbf{x}_{k0}}(\hat{\mathbf{x}}_1,t_{f,1})=0$ simultaneously for $t_{f,1}$ and $\hat{\mathbf{x}}_1$. We get analytical solution for ${t}_{f,1}$, $\hat{x}_1$ and $\dot{\hat{x}}_1$ using the Gr\"{o}bner basis based elimination \cite{cox2013ideals}. Since there are three equations and three unknowns $(\hat{x}_1,\dot{\hat{x}}_1, t_{f,1})$, we use lexicographic ordering $(\hat{x}_1\succ \dot{\hat{x}}_1 \succ t_{f,1})$ to arrive at the final expressions. The computation of the Gr\"{o}bner basis is performed using the Sagemath \cite{stein2007sage}.  
Subsequently, we get a quadratic equation in $b_2t_{f,1}^2+b_1t_{f,1}+b_0=0$.
Among both the solutions to the quadratic equation, we pick the one with the least real positive value as  $t_{f,1}$. Then, the corresponding intersection point $\hat{\mathbf{x}}_1$ is
\begin{align}\label{eq:st01}
\hat{x}_1&
=
\frac{1}{2}\hat{\dot{x}}\,t_{f,1}
+
\frac{1}{4}\!\left[
\hat{\dot{x}}\!\left(\dot{x}_{i0}-\dot{x}_{k0}\right)
+
t_{f,1}\!\left(\dot{x}_{i0}+\dot{x}_{k0}\right)
\right] \nonumber\\ &
+
\frac{1}{2}\!\left(x_{i0}+x_{k0}\right)
+
\frac{1}{8}\!\left(\dot{x}_{k0}^{2}-\dot{x}_{i0}^{2}\right)\nonumber
\\ 
   \dot{\hat{{x}}}_1&=
\frac{t_{f,1}(\dot{x}_{j0}
- \dot{x}_{k0})
+ 2(x_{j0} - x_{k0})
+ 0.5(\dot{x}_{j0}^{2}
-\dot{x}_{k0}^{2})
}{\dot{x}_{j0}-\dot{x}_{k0}
}\nonumber
\end{align}
At the computed time $t_{f,1}$, these sets meet at a unique point $\hat{\mathbf{x}}_1$. An example of a unique intersection point with $\partial{\mathcal{A}}_{i}^{\beta,s_1} \cap \partial{\mathcal{A}}_{j}^{\beta,s_3} \cap \partial{\mathcal{A}}_{k}^{\beta,s_1} $ is shown in Fig. \ref{fig2}.

\subsubsection{ $\partial{\mathcal{A}}_{i}^{\beta,s_3} \cap \partial{\mathcal{A}}_{j}^{\beta,s_2} \cap \partial{\mathcal{A}}_{k}^{\beta,s_1}$ }\label{sec:boundaryequations2}

Using boundary equations \eqref{eq:bds3} and \eqref{eq:bds2}, we solve $\Gamma^{\beta,s_3}_{\mathbf{x}_{i0}}(\hat{\mathbf{x}}_7,t_{f,7}) = \Gamma^{\beta,s_2}_{\mathbf{x}_{j0}}(\hat{\mathbf{x}}_7,t_{f,7}) = \Gamma^{\beta,s_1}_{\mathbf{x}_{k0}}(\hat{\mathbf{x}}_7,t_{f,7})=0$ for $t_{f,7}$ and $\hat{\mathbf{x}}_7$.
The expression for $t_{f,7}$ and the corresponding states $\hat{\mathbf{x}}_7$ obtained by solving these three equations are given below:
{\small
\begin{equation*}\label{eq:tf07}
t_{f,7}
=
\frac{
-(\beta+\dot{x}_{j0})\!\left(\dot{x}_{i0}+\dot{x}_{j0}+\dot{x}_{k0}+\beta\right)
+2\!\left(x_{i0}-x_{k0}\right)
+\dot{x}_{i0}^{2}
}{
2\beta-\dot{x}_{i0}+\dot{x}_{k0}
}
\end{equation*}
}
{\small\begin{equation}
\begin{aligned} \label{eq:st07}
\hat{x}_7 &=
\frac{1}{2}t_{f,7}\!\left(\dot{x}_{j0}+\dot{x}_{k0}+\beta\right)
+\frac{1}{4}\!\left(\dot{x}_{j0}-\dot{x}_{k0}+\beta\right)^{2}
-\frac{1}{4}\beta^{2}
-x_{k0} \\ 
   \dot{\hat{x}}_7 &= \dot{x}_{j0}+\beta \nonumber
 \end{aligned}
 \end{equation}
} 

The expressions of $t_{f,m}$ and $\hat{\mathbf{x}}_m$ for the remaining possibilities can be obtained similarly. 

For any agent $a_\alpha,~\alpha\in\{i,j,k\}$, if the boundary $\partial\mathcal{A}_\alpha^{\beta,s_l}$ is involved in the intersection, the switching times $t_{1\alpha}^{m,s_l}$ and $t_{2\alpha}^{m,s_l}$ are given by \eqref{eq:switchtimess1}, \eqref{eq:switchtimess2}, \eqref{eq:switchtimess3}, and \eqref{eq:switchtimess4} for $l=1,2,3,4$ respectively. Finally, we pick the value of $m$, denoted by $\bar{m}$, for which the inequality $0 \leq t^{m,\bullet}_{1 \alpha} \leq t^{m,\bullet}_{2 \alpha} \leq t_{f,m} < \infty$ holds for all $\alpha \in \{ i,j,k\}$, and $t_{f,m}$ is the \emph{least}. 
{
\begin{proposition}
Suppose the condition in {Case (ii)} is true for a triplet of agents $a_i,a_j,a_k$. Then, the \emph{minimum time} to consensus is $\bar{t}_{ijk} = t_{f,\bar{m}}$ and the corresponding minimum time consensus point is $\bar{\mathbf{x}}_{ijk}=\hat{\mathbf{x}}_{\bar{m}}$.
\end{proposition}}
\subsection{\label{sec:min.timeforNagent}Minimum time consensus for $N$ agents}
Assume that the condition (i) of Theorem \ref{thm:feas}, $\dot{x}_{\max}-\dot{x}_{\min}< 2\beta$ holds for the initial conditions of all $N$ agents. Then, we first prove the following result.
\begin{lemma}\label{intersection}
If for a triplet of agents $\{a_i,a_j,a_k\}$ there exists a time ${t}_f>0$ at which $\mathcal{A}_i^\beta({t}_f, \mathbf{x}_{i0}) \cap \mathcal{A}_j^\beta({t}_f, \mathbf{x}_{j0}) \cap \mathcal{A}_k^\beta({t}_f, \mathbf{x}_{k0}) \neq \emptyset$ then for all time $t_f'>{t}_f$ the intersection of their attainable sets remains non empty i.e., $\mathcal{A}_i^\beta(t_f', \mathbf{x}_{i0}) \cap \mathcal{A}_j^\beta(t_f', \mathbf{x}_{j0}) \cap \mathcal{A}_k^\beta(t_f', \mathbf{x}_{k0}) \neq \emptyset.$  
\end{lemma}
{ {\begin{proof}
Consider a point $\hat{\mathbf{x}}\in \mathcal{A}_i^\beta({t}_f, \mathbf{x}_{i0}) \cap \mathcal{A}_j^\beta({t}_f, \mathbf{x}_{j0}) \cap \mathcal{A}_k^\beta({t}_f, \mathbf{x}_{k0})$.  By Lemma \ref{lem:inclusion},  $e^{A(t_f'-t_f)}\hat{\mathbf{x}}\in \mathcal{A}_i^\beta({t}_f', \mathbf{x}_{i0}) \cap \mathcal{A}_j^\beta({t}_f', \mathbf{x}_{j0}) \cap \mathcal{A}_k^\beta({t}_f', \mathbf{x}_{k0})$ and hence $\mathcal{A}_i^\beta({t}_f', \mathbf{x}_{i0}) \cap \mathcal{A}_j^\beta({t}_f', \mathbf{x}_{j0}) \cap \mathcal{A}_k^\beta({t}_f', \mathbf{x}_{k0})\ne \emptyset$.
\end{proof}}}
Helly's theorem and Lemma \ref{intersection} give us the following Theorem. 

\begin{theorem}\label{thm:n-consensus}
    For $N$ agents the minimum time to consensus $\bar{t}_f = \max\limits_{1 \leq i,j,k \leq N}\bar{t}_{ijk}.$ 
\end{theorem}
 { \begin{proof}
 Let $\bar{t}_f = \max\limits_{1 \leq i,j,k \leq N}\bar{t}_{ijk}$ and the agents $(a_p,a_q,a_r)$ be such that  $(p,q,r)=\arg\max\limits_{1 \leq i,j,k \leq N}\bar{t}_{ijk}$.  For all other triplets $(a_i,a_j,a_k)$ of agents $\bar{t}_{ijk}\le \bar{t}_{pqr}$. Lemma \ref{intersection} ensures that $\mathcal{A}_i^{\beta}(\bar{t}_{pqr}, \mathbf{x}_{i0}) \cap \mathcal{A}_j^{\beta}(\bar{t}_{pqr}, \mathbf{x}_{j0}) \cap \mathcal{A}_k^{\beta}(\bar{t}_{pqr}, \mathbf{x}_{k0}) \neq \emptyset$ for all the triplets $(a_i,a_j,a_k)$. Thus, using Helly's theorem at time $\bar{t}_{pqr}$ we have $\bigcap_{i=1}^N \mathcal{A}_i^{\beta}(\bar{t}_{pqr},\mathbf{x}_{i0})\ne \emptyset$. Further, for any time $t_f<\bar{t}_{pqr}$,  $\mathcal{A}_p^{\beta}({t}_{f}, \mathbf{x}_{p0}) \cap \mathcal{A}_q^{\beta}({t}_{f}, \mathbf{x}_{q0}) \cap \mathcal{A}_r^{\beta}({t}_{f}, \mathbf{x}_{r0}) = \emptyset$ and hence, $\bigcap_{i=1}^N \mathcal{A}_i^{\beta}({t}_{f},\mathbf{x}_{i0})= \emptyset$. Therefore, $\bar{t}_{pqr}= \max\limits_{1 \leq i,j,k \leq N}\bar{t}_{ijk}$ is the minimum time at which   $\bigcap_{i=1}^N \mathcal{A}_i^{\beta}(\bar{t}_{f},\mathbf{x}_{i0})\ne \emptyset$. Thus, the minimum time to consensus is $\bar{t}_f = \max\limits_{1 \leq i,j,k \leq N}\bar{t}_{ijk}$.
 \end{proof}}
 The Theorem \ref{thm:n-consensus} allows us to obtain $\bar{t}_f$ and $\bar{\mathbf{x}}$ by combining the minimum time to consensus obtained for all the $\Mycomb[N]{3}$ triplets, which allows the computation to be distributed among agents. {To this end, we require the following assumption.
 \begin{assumption}
    All agents communicate over a connected graph.
\end{assumption} }
 \subsection{Distributed Computation}
{ The common information known to all agents is the number of agents $N$, agent ids $a_i$ where $i=1,...,N$, and an ordered set of $\Mycomb[N]{3}$ triplet of agents $\mathscr{T}:=\{\nu_1,...\nu_{\Mycomb[N]{3}}\}$. The following steps are performed for implementing the computation on board each agent:
\begin{enumerate}
\item[1)] Each agent maintains a list with the agent id and corresponding initial condition. This list is updated by sharing information between connected agents. There are several distributed routing-based broadcasting algorithms  \cite[Chapter 5]{datanetworks} for this purpose that can be implemented using mobile ad hoc networks. 
\begin{algorithm} 
\caption{\label{alg:tripassign} Computation assigned to agent $a_i$}
\begin{algorithmic}[1]
\State \textbf{Inputs:} $N$, $\mathbf{x}_{j0}$ for $j=1,...,N$, $\mathscr{T}:=\{\nu_1,...\nu_{\Mycomb[N]{3}}\}$ 
\State \textbf{Outputs:} Set $\mathscr{T}_i$ of triplets assigned to agent $a_i$, $\nu_{\max,i}$, $t_{\nu_{\max,i}}$ and $\bar{\mathbf{x}}_{\nu_{\max,i}}$ 
\State $\sigma:=\lfloor\frac{1}{N}\Mycomb[N]{3}\rfloor$ and $\rho := \Mycomb[N]{3} \mod{N} $ 
\If {$i\le \rho$}
       \State Assign the triplets $\mathscr{T}_i:=\{\nu_{(i-1)\sigma+1},...,\nu_{i\sigma},\nu_{N\sigma+i}\}$ to agent $a_i$. 
\Else
        \State  Assign the triplets $\mathscr{T}_i:=\{\nu_{(i-1)\sigma+1},...,\nu_{i\sigma}\}$ to agent $a_i$
\EndIf
\State Using procedure developed in Section \ref{min.time.for.N=3}, compute  $\bar{t}_{\nu}$ and $\bar{\mathbf{x}}_{\nu}$ for all triplets $\nu\in\mathscr{T}_i$.

\State $\nu_{\max,i}=\arg\max\limits_{\nu\in \mathscr{T}_i} \bar{t}_{\nu}$ 
\State $t_{\nu_{\max,i}}=\max\limits_{\nu\in \mathscr{T}_i} \bar{t}_{\nu}$ and  $\bar{\mathbf{x}}_{\nu_{\max,i}}$ of the triplet $\nu_{\max,i}$.
\end{algorithmic}
\end{algorithm}
\item[2)] According to the 
Algorithm \ref{alg:tripassign}, each agent $a_i$ gets assigned at most $\sigma+1$ number of triplets for computation. Agent $a
_i$ computes and shares $\nu_{\max,i}$, $t_{\nu_{\max,i}}$ and $\bar{\mathbf{x}}_{\nu_{\max,i}}$. 
\item[3)] {The sharing of the information $\nu_{\max,i}$, $t_{\nu_{\max,i}}$ and $\bar{\mathbf{x}}_{\nu_{\max,i}}$by each agent $a_i$ is again performed by using distributed routing as done in step 1. By using this step all agents receive $\nu_{\max,w}$, $t_{\nu_{\max,w}}$ and $\bar{\mathbf{x}}_{\nu_{\max,w}}$ computed by each agent $a_w$, $w=1,2,...,N$.}
\item[4)]  {{Note that as per Theorem \ref{thm:n-consensus},  $\bar{t}_f=\max\limits_{\nu\in\mathcal{T}}\bar{t}_\nu
=\max\limits_{w=1,\dots,N}\ \max\limits_{\nu\in\mathcal{T}_w}\bar{t}_\nu
=\max\limits_{w=1,\dots,N}t_{\nu_{\max,w}}$.} Thus, each agent locally implements Algorithm \ref{alg:inputs} to compute $\bar{t}_f=\max\limits_{w=1,...,N}{t_{\nu_{\max,w}}}$ as the minimum time to consensus. Let $w^*:=\arg\max\limits_{w=1,...,N}{t_{\nu_{\max,w}}}$. Then the corresponding consensus point is computed to be $\bar{\mathbf{x}} = \hat{\mathbf{x}}_{\nu_{\max,w^*}}$. }
\item[5)] {Finally, in same Algorithm \ref{alg:inputs}, each agent $a_i$ also computes the input $u_i(t)$ of the form \eqref{twoswitch} required to attain the optimal consensus point  $\bar{\mathbf{x}}$ exactly at time $\bar{t}_f$ and implements it onboard. }
\end{enumerate}

\begin{algorithm} {
\caption{\label{alg:inputs} Local computation of control input for agent $a_i$}

\begin{algorithmic}[1]
\State \textbf{Inputs:} For $w=1,..,N$, the triplet $\nu_{\max,w}$, $t_{\nu_{\max,w}}$ and $\bar{\mathbf{x}}_{\nu_{\max,w}}$ obtained from neighbours using distributed routing. 
\State \textbf{Outputs:} $\bar{t}_f$, $\bar{\mathbf{x}}$, and control inputs $u_i(t)$ of the form \eqref{twoswitch}. 
\State $w^*:=\arg\max_{w}{t_{\nu_{\max,w}}}$, $\bar{t}_f={t_{\nu_{\max,w^*}}}$, and $\bar{\mathbf{x}}=\hat{\mathbf{x}}_{\nu_{\max,w^*}}$ \\
For any $l=1,2,3,4$, \If { $\bar{\mathbf{x}} \in  {\partial\mathcal{A}}_i^{\beta,s_l}(\bar{t}_f,\mathbf{x}_{i0})$}
       \State Use input sequence $s_l$ with switching times $t_1$ and $t_2$ given by the equations \eqref{eq:switchtimess1} if $l=1$, \eqref{eq:switchtimess2} if $l=2$, \eqref{eq:switchtimess3} if $l=3$, and \eqref{eq:switchtimess4} if $l=4$. 
\Else{ $\bar{\mathbf{x}} \in  {\mbox{int }\mathcal{A}}_i^{\beta}(\bar{t}_f,\mathbf{x}_{i0})$}  
\State Compute $0<\hat{\beta}<\beta$ s.t., $\bar{\mathbf{x}} \in  {\partial\mathcal{A}}_i^{\hat{\beta}}(\bar{t}_f,\mathbf{x}_{i0})$ 

\State For any $l=1,2,3,4$, \If{ $\bar{\mathbf{x}} \in  {\partial\mathcal{A}}_i^{\hat{\beta},s_l}(\bar{t}_f,\mathbf{x}_{i0})$}
       \State Use input sequence $s_l$ with switching times $t_1$ and $t_2$ given by the equations \eqref{eq:switchtimess1} if $l=1$, \eqref{eq:switchtimess2} if $l=2$, \eqref{eq:switchtimess3} if $l=3$, and \eqref{eq:switchtimess4} if $l=4$.
\EndIf 
\EndIf
\end{algorithmic}}
\end{algorithm}
}

{In Algorithm \ref{alg:inputs}, in the case where $\bar{\mathbf{x}} \in \mbox{int } {\mathcal{A}}_i^\beta(\bar{t}_f,\mathbf{x}_{i0})$ holds, the final consensus state $\bar{\mathbf{x}}$ can be attained from $\mathbf{x}_{i0}$ at $\bar{t}_f$ using strictly lesser amount of fuel than $\beta$. Therefore, we recompute the lesser fuel requirement $\hat{\beta}\in(0,\beta)$ for such agents. This is done by solving equations $\Gamma^{{\beta_{s_1}},s_1}_{\mathbf{x}_{i0}}(\bar{\mathbf{x}},\bar{t}_f)=0$,  $\Gamma^{{\beta_{s_3}},s_3}_{\mathbf{x}_{i0}}(\bar{\mathbf{x}},\bar{t}_f)=0$, $\Gamma^{{\beta_{s_2}},s_2}_{\mathbf{x}_{i0}}(\bar{\mathbf{x}},\bar{t}_f)=0$,  or $\Gamma^{{\beta_{s_4}},s_4}_{\mathbf{x}_{i0}}(\bar{\mathbf{x}},\bar{t}_f)=0$ for $\beta_{s_l}, l=1,2,3,4$ respectively. Since $\bar{\mathbf{x}}$ and $\bar{t}_f$ are known, each of these equations are quadratic in ${\beta_{s_l}}$, $l=1,2,3,4$ respectively. Once $\beta_{s_l}$ is obtained, we also compute the switching times $t_1^{s_l}$ and $t_2^{s_l}$ given by \eqref{eq:switchtimess1}, \eqref{eq:switchtimess2}, \eqref{eq:switchtimess3}, and \eqref{eq:switchtimess4} for $l=1,2,3,4$ respectively. Among all the solutions $\beta_{s_l}$, the least real positive valued solution which lies in the range $(0,\beta)$ and satisfying the inequality $0\le t_1^{s_l}\le t_2^{s_l}\le t_f$ is chosen as $\hat{\beta}$. }
\subsubsection{{Computational Complexity}}\label{complexity}
The computation for a particular triplet takes a finite and fixed number of flops. For $N$-agents, there are $\Mycomb[N]{3}= \frac{N(N-1)(N-2)}{6}$
such triplets to be considered in the computation. Thus, the entire computation requires only $O(N^3)$ flops (as opposed to $O(2^N)$ steps required when directly computing without using Helly's theorem).
Furthermore, each triplet can be handled independently; thus, it is possible to distribute the computation (see Algorithm \ref{alg:tripassign}) so that each agent handles only $O(N^2)$ flops. 

{Note that the initial conditions are in the reachable set, i.e., $\mathbf{x}_{j0}\in \mathcal{R}^{\beta}(\bar{t}_f,\bar{\mathbf{x}}),~j=1,...,N$. Since $\mathcal{R}^{\beta}(\bar{t}_f,\bar{\mathbf{x}})$ is convex, the convex hull of all initial conditions $\mathbf{x}_{j0}$, $j=1,..,N$ is contained in $\mathcal{R}^{\beta}(\bar{t}_f,\bar{\mathbf{x}})$. Thus, $\bar{t}_f$, and $\bar{\mathbf{x}}$ are determined by only those agents whose initial conditions are on the $\partial\mathcal{R}^{\beta}(\bar{t}_f,\bar{\mathbf{x}})$. As a result, while computing $\bar{t}_f$ and $\bar{\mathbf{x}}$, only those triplets having agents with initial conditions that are on the extreme points of the convex hull have to be considered. This reduces the computational effort significantly.}
\begin{figure}
\centering
\includegraphics[scale=0.28]{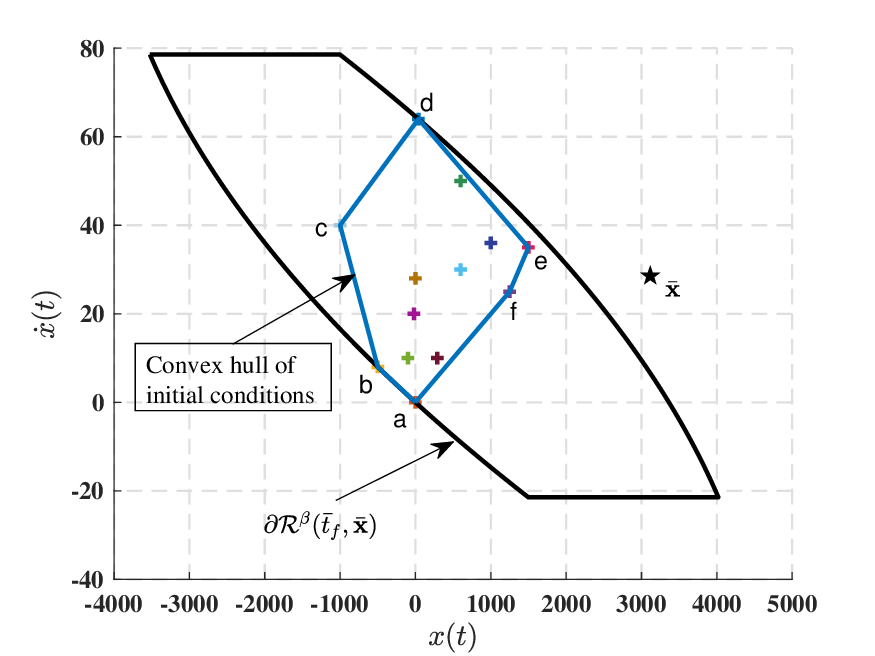}
\caption{\label{fig:reachconvhull}{{Convex hull of initial conditions inside reachable set}}}
\end{figure}
For example, in Fig. \ref{fig:reachconvhull}, the convex hull of all initial conditions is the polygon $abcdef$, which is contained in the reachable set. Thus, only the triplets formed by the agents with initial conditions at the extreme points $a,b,c,d,e$ and $f$ determine the $\bar{t}_f$ and $\bar{\mathbf{x}}$.
\section{\label{sec:example}Example} 
{Consider a set of initial conditions for twelve double integrator agents, $a_1,...,a_{12}$ i.e.,  $\mathbf{x}_{10} =[0,0]^{\top}$, $\mathbf{x}_{20} =[40,64]^{\top}$, $\mathbf{x}_{30} =[-500,8]^{\top}$, $\mathbf{x}_{40} =[-100,10]^{\top}$, $\mathbf{x}_{50} =[600,30]^{\top}$, $\mathbf{x}_{60} =[2900,10]^{\top}$, $\mathbf{x}_{70} =[0,  28]^{\top}$, $\mathbf{x}_{80} =[1250, 25]^{\top}$, $\mathbf{x}_{90} =[1500, 35]^{\top}$, $\mathbf{x}_{100} =[ 600, 50]^{\top}$, $\mathbf{x}_{110} =[1000, 36]^{\top}$, $\mathbf{x}_{120} =[-20, 20]^{\top}$ respectively. Note that $\dot{x}_{\max}=64$ and $\dot{x}_{\min}=0$.  Let the fuel budget of each agent be $\beta = 50$. Thus, the set of initial conditions satisfies the inequality $\dot{x}_{\max}-\dot{x}_{\min}<2\beta$. Using Lemma \ref{lemma:consensuspossible}, we conclude that the minimum time consensus point $\bar{\mathbf{x}}\in\mathcal{X}^{\beta} = \{(x,\dot{x})\in \mathbb{R}^2| \dot{\bar{x}}\in [14,50]\}$.  For computing $\bar{\mathbf{x}}$ and $\bar{t}_f$, we perform minimum time consensus computation as per the steps given in Section \ref{min.time.for.N=3} for each of $\Mycomb[12]{3}=220$ the triplets of agents. From Theorem \ref{thm:n-consensus}, the minimum time to consensus is obtained from the triplet of agents $\{a_1,a_2,a_3\}$. The minimum time to consensus is $\bar{t}_f=100.44$ and the corresponding consensus point is   $\bar{\mathbf{x}}=\begin{bmatrix} 3116.36 & 28.56 \end{bmatrix}^{\top}\in \mathcal{X}_c^{50}$. For agents $a_i\in\{a_1,a_2,a_3\}$,  $\bar{\mathbf{x}} \in \partial \mathcal{A}_i^\beta(\bar{t}_f,\mathbf{x}_{i0})$. For the remaining agents $a_i\in \{a_4,a_5,a_6,a_7,a_8,a_9,a_{10},a_{11},a_{12}\}$, the optimal consensus point $\bar{\mathbf{x}} \in \text{int } \mathcal{A}_i^\beta(\bar{t}_f,\mathbf{x}_{i0})$. To reach the consensus point exactly at time $\bar{t}_f$, inputs are computed using Algorithm \ref{alg:inputs}. The control inputs for all agents are listed in Table \ref{tab1}. In Fig. \ref{fig:3b} and \ref{fig:3c}, we plot the state-trajectories and the attainable sets for all agents, respectively. Further, the same results are obtained by considering four agents $a_1,a_2,a_3,$ and $a_6$ that are the extreme points of the convex hull, thereby reducing the computation significantly to only four triplets.} 
\begin{figure}
    \centering
    \begin{subfigure}[t]{0.48\linewidth}
        \centering   \includegraphics[width=\linewidth,height=3.5cm]{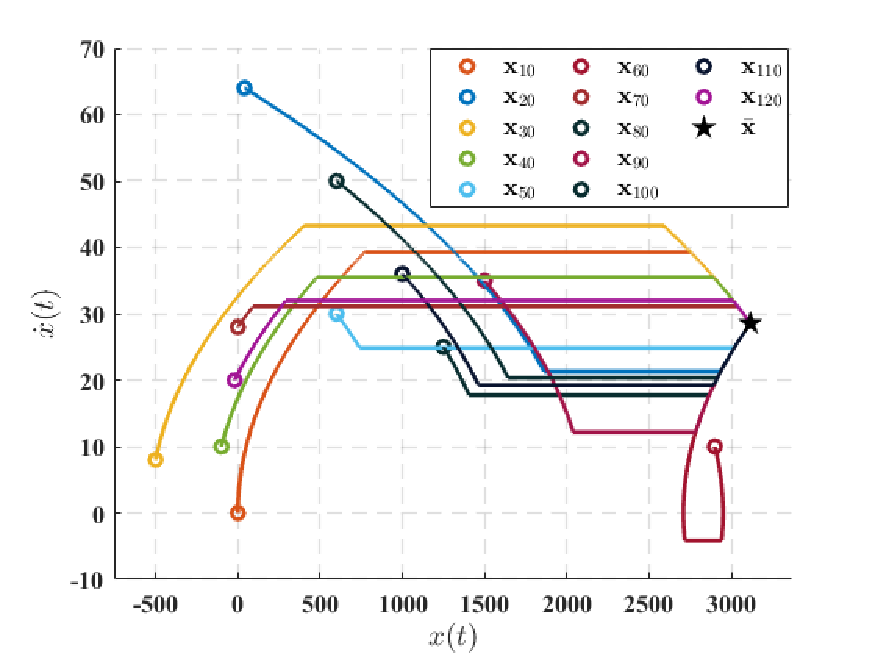}
        \caption{\small{Phase plots}}
        \label{fig:3b}
    \end{subfigure}
    \hfill
    \begin{subfigure}[t]{0.48\linewidth}
        \centering     \includegraphics[width=1.1\linewidth,height=3.5cm]{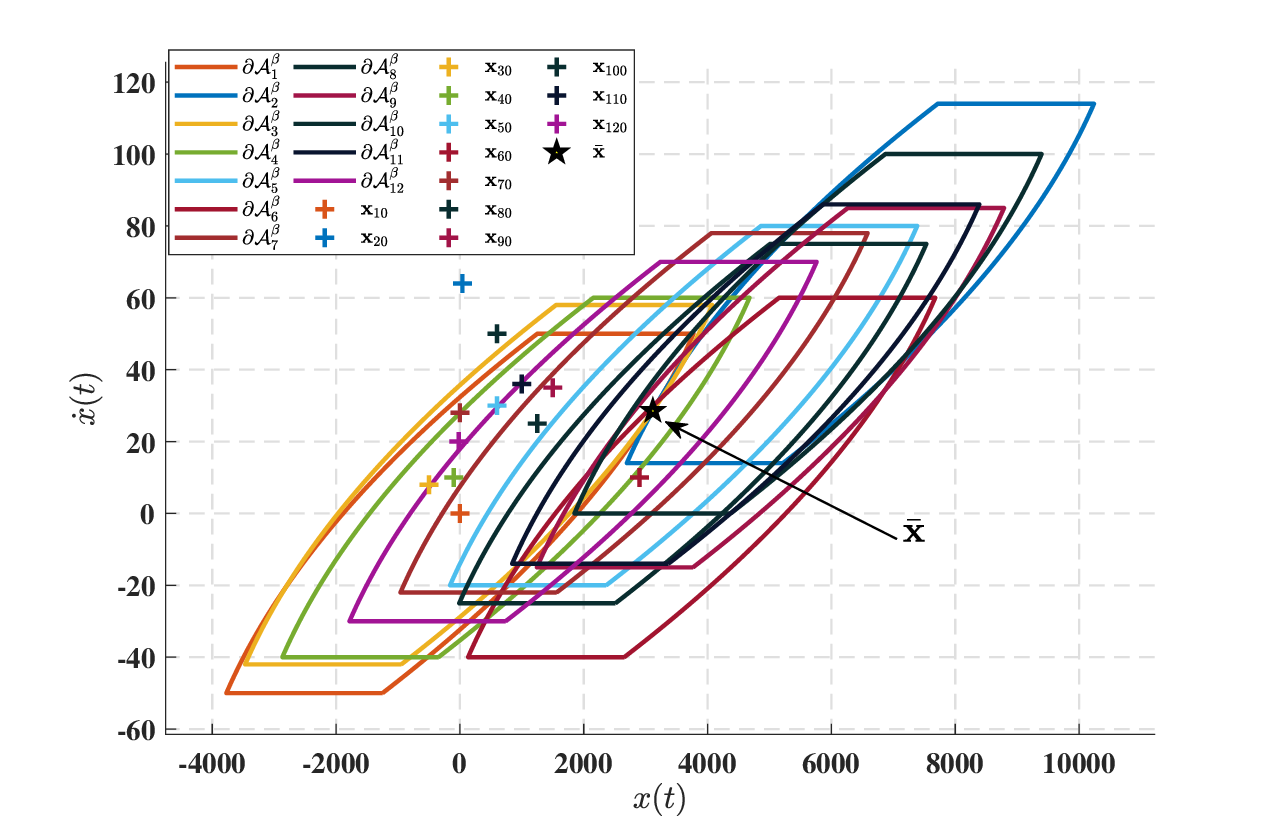}    \caption{\small{Attainable set intersection}}
        \label{fig:3c}
    \end{subfigure}
    \caption{{State trajectories and attainable sets for $N=12$ agents.}}
    \label{Fig:3}
\end{figure}

\begin{table}
\setlength{\abovecaptionskip}{2pt}
\setlength{\belowcaptionskip}{0pt}
\caption{\label{tab1}{\small Fuel $\beta_i$, switching times $(t_1,t_2)$, and optimal control
$u_i^\ast$.}}

\centering\footnotesize
\setlength{\tabcolsep}{3.5pt}
\renewcommand{\arraystretch}{1.0}
\begin{tabular}{@{}lrrrc@{\hspace{0.9em}}|@{\hspace{0.9em}}lrrrc@{}}
\hline
$a_i$ & $\beta_i$ & $t_1$ & $t_2$ & $u_i^\ast$ &
$a_i$ & $\beta_i$ & $t_1$ & $t_2$ & $u_i^\ast$ \\
\hline
$a_1$ & 50.00 & 39.28 & 89.72 & $s_1$ & $a_7$    &  5.65 &  3.10 & 97.89 & $s_1$ \\
$a_2$ & 50.00 & 42.72 & 93.16 & $s_3$ & $a_8$    & 18.08 &  7.26 & 89.62 & $s_3$ \\
$a_3$ & 50.00 & 35.28 & 85.72 & $s_1$ & $a_9$    & 39.25 & 22.84 & 84.03 & $s_3$ \\
$a_4$ & 32.44 & 25.50 & 93.50 & $s_1$ & $a_{10}$ & 37.88 & 29.66 & 92.22 & $s_3$ \\
$a_5$ &  8.86 &  5.15 & 96.73 & $s_3$ & $a_{11}$ & 26.08 & 16.76 & 91.12 & $s_3$ \\
$a_6$ & 46.94 & 14.19 & 67.69 & $s_3$ & $a_{12}$ & 15.45 & 12.00 & 96.99 & $s_1$ \\
\hline
\end{tabular}
\end{table}

\section{\label{sec5}Conclusions}
For $N$ double-integrator agents, we computed the minimum time to consensus and
the corresponding consensus point under bounded inputs and a fixed fuel budget.
The key tool is the attainable set under fuel constraints, which we characterized
along with its properties. Convexity breaks the $N$-agent problem into multiple three-agent problems via Helly's theorem, distributing the computation across agents.
We also characterized the initial configurations from which consensus is possible.
The approach extends to any system with convex attainable sets and tractable
boundaries, including higher-order linear systems. Future work will pursue a
receding-horizon closed-loop scheme in which $\bar{\mathbf{x}}$ and $\bar{t}_f$
are recomputed periodically, and extensions to nonlinear dynamics.

\section*{Acknowledgment}
The authors thank anonymous reviewers for their valuable suggestions that helped in improving this technical note. 
\section*{References}
\bibliographystyle{IEEEtran}
\bibliography{reference} 

\end{document}